\documentclass[5p,times]{elsarticle}

\usepackage{lineno,hyperref}
\modulolinenumbers[5]
\usepackage[dvipsnames]{xcolor}
\usepackage{algpseudocode}
\usepackage{algorithm}
\usepackage{graphicx}
\usepackage{adjustbox}
\usepackage{amsmath,amssymb} 
\usepackage{tabularx}
\usepackage{multirow}
\usepackage{tablefootnote}
\usepackage{subcaption}
\usepackage{soul,color}
\usepackage{float}
\usepackage{graphicx,wrapfig,picins}

\usepackage{dblfloatfix}
\usepackage{float}
\usepackage{bbding}
\usepackage{lineno,hyperref}
\journal{Computers \&  Security}

\begin{document}

\begin{frontmatter}

\title{Adv-Bot: Realistic Adversarial Botnet Attacks \\ against Network Intrusion Detection Systems}

\author[mymainaddress,mysecondaryaddress]{Islam Debicha\corref{mycorrespondingauthor}}
\cortext[mycorrespondingauthor]{Corresponding author}
\ead{islam.debicha@ulb.be}

\author[mymainaddress]{Benjamin Cochez\corref{mycorrespondingauthor}}
\ead{benjamin.cochez@ulb.be}
\author[mythirdaddress]{Tayeb Kenaza}
\author[mysecondaryaddress]{Thibault Debatty}
\author[mymainaddress]{Jean-Michel Dricot}
\author[mysecondaryaddress]{Wim Mees} 

\address[mymainaddress]{ Cybersecurity Research Center, Université Libre de Bruxelles, 1000 Brussels, Belgium}
\address[mysecondaryaddress]{Cyber Defence Lab, Royal Military Academy, 1000 Brussels, Belgium}
\address[mythirdaddress]{Computer Security Laboratory, Ecole Militaire Polytechnique, Algiers, Algeria}

\begin{abstract}

Due to the numerous advantages of machine learning (ML) algorithms, many applications now incorporate them. However, many studies in the field of image classification have shown that MLs can be fooled by a variety of adversarial attacks. These attacks take advantage of ML algorithms' inherent vulnerability. This raises many questions in the cybersecurity field, where a growing number of researchers are recently investigating the feasibility of such attacks against machine learning-based security systems, such as intrusion detection systems. The majority of this research demonstrates that it is possible to fool a model using features extracted from a raw data source, but it does not take into account the real implementation of such attacks, i.e., the reverse transformation from theory to practice. The real implementation of these adversarial attacks would be influenced by various constraints that would make their execution more difficult. As a result, the purpose of this study was to investigate the actual feasibility of adversarial attacks, specifically evasion attacks, against network-based intrusion detection systems (NIDS), demonstrating that it is entirely possible to fool these ML-based IDSs using our proposed adversarial algorithm while assuming as many constraints as possible in a black-box setting. In addition, since it is critical to design defense mechanisms to protect ML-based IDSs against such attacks, a defensive scheme is presented. Realistic botnet traffic traces are used to assess this work. Our goal is to create adversarial botnet traffic that can avoid detection while still performing all of its intended malicious functionality.

\end{abstract}


\end{frontmatter}


\section{Introduction}

Sophisticated methods of cybercrime, such as botnets, are becoming increasingly rampant. Given the severity of the threat, it is essential to have a defense mechanism that can detect all types of botnet traffic. Among these mechanisms, the use of intrusion detection systems dedicated to network analysis, known as NIDS, is gaining popularity. And given the difficulty of some signature-based NIDSs in detecting new botnet attacks or even variants of known botnet attacks, NIDSs based on machine learning algorithms have become more prevalent. These machine learning-based NIDS can detect not only variants of known attacks, but also novel attacks, also known as zero-day attacks \cite{DBLP:journals/access/XinK0CLZGHW18,DBLP:journals/ijon/MahdavifarG19}.

Despite this encouraging achievement, many recent studies have shown that it is possible to fool the ML algorithms used in these detection methods \cite{apruzzese2019evaluating,hashemi2019towards}, as has been discovered previously in other application areas, such as computer vision \cite{DBLP:journals/corr/SzegedyZSBEGF13,kurakin2018adversarial}. These ML algorithms have been shown to be vulnerable to a variety of adversarial attacks, including poisoning, extraction, and evasion. Evasion attacks are studied in this paper because they are more realistic in cyber security scenarios than the other two types of attacks \cite{apruzzese2022modeling}. Evasion attacks can fool any ML model during its inference process by adding slight, often imperceptible, perturbations to the original instance sent to create so-called adversarial instances.


The literature study shows that there has been considerable research on the impact of adversarial attacks on machine learning models \cite{miller2020adversarial,zhang2019adversarial,qiu2019review}. However, their feasibility in domain-constrained applications, such as intrusion detection systems, is still in its early stages \cite{martins2020adversarial,vitorino2022adaptative,mccarthy2022functionality}. Adversarial attacks can be performed in either white-box or black-box settings. Many white-box adversarial attacks, originally developed for computer vision applications \cite{DBLP:journals/corr/GoodfellowSS14,DBLP:conf/iclr/MadryMSTV18,DBLP:conf/sp/Carlini017}, have been applied directly to network traffic without addressing domain constraints properly \cite{DBLP:journals/fgcs/PawlickiCK20,debicha2023tad, merzouk2022investigating}.

Aside from the issue of functionality preservation, white-box attacks necessitate knowledge of the target system's internal architecture. It is unlikely that an attacker would have access to the internal configuration of the ML model \cite{miller2020adversarial}, making a white-box attack in a realistic environment less likely \cite{apruzzese2022modeling}. As a result, recent research \cite{apruzzese2020deep, zhang2022adversarial} has focused on designing adversarial network traffic in a black-box setting where the attacker has little or no knowledge of the defender's NIDS. Model querying \cite{chen2017zoo} and transferability \cite{DBLP:journals/corr/GoodfellowSS14} are two methods for launching black-box attacks.


An attacker can extract useful information such as the classifier's decision by sending multiple queries to the target NIDS, allowing the attacker to craft adversarial perturbations capable of evading the NIDS \cite{apruzzese2020deep,zhang2019evasion,ren2020query}. Although this approach achieves high success rates, it has two limitations. The first is that NIDSs are not designed to provide feedback when queried, unless side-channel attacks are used \cite{boenisch2021side}. The second limitation is that it requires a relatively high number of queries to function properly, exposing the attacker to detection by a simple query detector \cite{zhang2020tiki, zhang2022adversarial}. As a result, for a realistic attack, the IDS querying approach is less feasible \cite{apruzzese2022modeling}.


The transferability property of adversarial examples is used as an alternative black-box approach. Goodfellow et al. \cite{DBLP:journals/corr/GoodfellowSS14} were the first to investigate this property in computer vision, demonstrating that adversarial examples that fool one model can fool other models with a high probability without the need for the models to have the same architecture or be trained on the same dataset. An attacker can use the transferability property to launch black-box attacks by training a surrogate model on the same data distribution as the target model \cite{debicha2021detect}. Sniffing network traffic is a simple way to accomplish this, especially in a botnet scenario where the attacker already has a foothold in the corporate network \cite{al2020real}. To the best of our knowledge, this paper represents the first proposal that exploits the transferability property to design a realistic black-box evasion attack by considering the constraints of the NIDS domain to create adversarial botnet traffic.


This paper provides a dedicated framework for leveraging evasion attacks to mislead the NIDS into classifying botnet traffic as benign under realistic constraints. Three contributions are included in this work: 
\begin{itemize}
    \item An in-depth analysis of the feasibility constraints required to generate valid adversarial perturbations while preserving the underlying logic of the network attack.
    \item A new black-box adversarial algorithm that can generate valid adversarial botnet traffic capable of evading botnet detection without any knowledge of the target NIDS.
    \item A defense that allows the proposed botnet evasion attack to be mitigated. This defense is inspired by adversarial detection and an ensemble method known as bagging.
\end{itemize}  

The remainder of the paper is organized as follows. Background and related work are presented in Section \ref{sec:back}. The proposed method is explained in Section \ref{sec:proposed}. Results and discussions are presented in Section \ref{sec:results}. Concluding remarks and suggestions for possible follow-up work are given in Section \ref{conclusion}.

\section{Background \& Related work}
\label{sec:back}
\subsection{Background}
\label{subsec:criteria}
With the increasing research on evasion attacks, the feasibility of such attacks in the real world is gaining more and more attention, regardless of the application domain. It is possible to distinguish three criteria that influence the realism of such attacks, namely: knowledge restriction, domain constraints, and manipulation space. The focus of this paper is on the analysis of these criteria in the domain of network-based intrusion detection system. 

\subsubsection{Knowledge Restriction}
Adversarial attacks come in three varieties: white-box, grey-box, and black-box. White-box attacks mean that the adversary knows everything about the model architecture and training dataset, in particular all the parameters and meta-parameters, which are, for example, the inputs and gradients in the case of neural networks, or the tree depth for decision trees. In addition, the adversary may also know the chosen cost function or the optimizer type in the case of neural networks. The gray-box attack, on the other hand, implies that the attacker has some limited knowledge of the target model. He may, for instance, know what data was used to train the model without knowing the type of model used or its internal architecture. A black-box attack occurs when the attacker does not know the architecture of the target model and the dataset used. Nevertheless, even without knowing anything about the model, it could still approach a decision boundary similar to that of the target model and thus fool it by querying the target model and receiving responses in the form of decisions or probabilities. An alternative black-box approach is known as "transferability" where the attacker can create his own model (i.e. surrogate model) with similar functionality to the target model in order to fool it by creating adversarial instances based on his surrogate model and then transferring these instances to the target model to fool it as well. Obviously, black-box attacks are more complicated to perform, not only due to lack of knowledge but also because they require more computational resources to accommodate these accumulated knowledge biases.

\subsubsection{Domain constraints}
Regarding the feasibility of adversarial attacks, it varies according to the domain in question as it is strongly limited by several constraints. Such constraints can be divided into two main categories: syntactic constraints and semantic constraints.

Syntactic constraints refer to all constraints related to syntax. In their work, Merzouk et al. \cite{merzouk2022investigating} identified three syntactic constraints that an adversarial instance must respect, namely out-of-range values, non-binary values, and multi-category membership. Out-of-range values are values that exceed a theoretical maximum value that cannot be exceeded. Non-binary values are entries that violate the binary nature of a feature, and multi-category membership are values that violate the concept of one-hot encoding.

Semantic links represent the connections that various features may have with each other. It is challenging to identify precisely such constraints since they are specific to each application and to each particular feature used. Nevertheless, the work of Hashemi et al. \cite{hashemi2019towards}, and Teuffeunbach et al. \cite{teuffenbach2020subverting} suggest an intuitive approach in the NIDS domain by categorizing the features into roughly three different groups with different semantic links. The first group includes features that can be directly modified by the attacker (e.g., the number of forwarding packets, the size of the forwarding packets, and flow duration). The second group concerns features that depend on the first group. They must be recalculated on the basis of the latter (e.g., number of packets/second or average forward packet size). The third group includes features that cannot be changed by the attacker (e.g., IP address, protocol number).

\subsubsection{Manipulation Space}
An essential property of a realistic adversarial instance is the ability of an attacker to modify its characteristics. In theory, it is possible to directly modify the features of adversarial instances. However, in real-world scenarios, this approach is considered unsuitable for certain domains, such as IDSs that analyze network traffic. This is mainly due to the fact that the feature extraction process (i.e., from raw traffic to feature space) is not a fully reversible process, unlike in other domains such as computer vision. This means that features can be extracted, modified, but not easily reintroduced into network traffic due to the semantic links between features. Moreover, direct feature modification requires full knowledge of the feature extraction process used by the IDS in order to respect the syntactic or semantic constraints assigned to them. We can therefore deduce that working on the feature space is not very realistic. For this reason, recent studies \cite{han2021evaluating,sadeghzadeh2021adversarial,chen2020generating} propose to directly manipulate the network traffic, so that it is not necessary to know the features used, nor to transform the feature values into traffic form. In this way, we can distinguish two manipulation spaces, the feature-based and the traffic-based.

\subsubsection{Performance Metrics}
All IDS models can be analyzed in terms of their performance. To evaluate their performance, different evaluation metrics can be used. All of them are based on the confusion matrix represented in Table \ref{tab:idsmetrics}.

\begin{table}[ht]
\caption{IDS Confusion matrix}
\begin{adjustbox}{width=0.8\columnwidth,center}
\begin{tabular}{|c|cc|}
\hline
                      & \multicolumn{2}{c|}{\textbf{Predicted Class}}                  \\ \hline
\textbf{Actual Class} & \multicolumn{1}{c|}{Anomaly}             & Normal              \\ \hline
Anomaly               & \multicolumn{1}{c|}{True Positive (TP)}  & False Negative (FN) \\ \hline
Normal                & \multicolumn{1}{c|}{False Positive (FP)} & True Negative (TN)  \\ \hline
\end{tabular}
\end{adjustbox}
\label{tab:idsmetrics}
\end{table}

Once all of the behavioral responses are classified, it is possible to determine evaluation metrics. Of these, the ones commonly used are :
\begin{enumerate}

        \item Recall, also called the "detection rate", is the percentage of the patterns classified as malicious and which are indeed malicious.
    \begin{equation}
        \label{eq:recall}
        \text{Recall} = \frac{TP}{(TP+FN)}
    \end{equation}
    \item Precision represents the percentage of relevant results instead of irrelevant results.
    \begin{equation}
        \label{eq:precision}
        \text{Precision} = \frac{TP}{(TP+FP)}
    \end{equation}
    \item F-measure, also called the "F1 score", gives the performance of the combined recall and precision evaluation metrics, it’s the harmonic mean of both. It provides the system with the capacity to give relevant results and refuse the others.
    \begin{equation}
        \label{eq:f1-score}
        \text{F-measure} = \frac{(2 * Recall * Precision)}{(Recall+Precision)}
    \end{equation}

\end{enumerate}

\subsection{Related work}
\label{sec:AnomalyDetection}

A common approach used by researchers in many areas is to focus on the theoretical aspect of the problem. In the case of adversarial machine learning, this entails focusing only on the feature level which is the ML representation of raw data after the feature extraction step \cite{wang2018deep,yang2018adversarial,martins2019analyzing}. Semantic and syntactic constraints, on the other hand, are two very important aspects to consider when transcribing a feature-represented instance into raw values in practice, and they are frequently overlooked or ignored. This implies that the realistic feasibility of adversarial attacks is not fully addressed \cite{apruzzese2022modeling}.


Merzouk et al. \cite{merzouk2022investigating} provide an in-depth analysis of the feasibility of some state-of-the-art white-box attacks against an IDS. The white-box adversarial algorithms studied are FGSM \cite{DBLP:journals/corr/GoodfellowSS14}, BIM \cite{kurakin2018adversarial}, C\&W \cite{DBLP:conf/sp/Carlini017}, DeepFool \cite{DBLP:conf/cvpr/Moosavi-Dezfooli16} and JSMA \cite{papernot2016limitations}. They have shown that these adversarial algorithms produce invalid adversarial instances and thus do not meet any of the feasibility criteria for a realistic adversarial attack as discussed in Section \ref{subsec:criteria}. In addition to the issue of functionality preservation, white-box attacks require knowledge of the internal architecture of the target system. It is unlikely that an attacker would have access to the internal configuration of the ML model \cite{miller2020adversarial} , making it less likely that an attacker could perform a white-box attack in a realistic environment \cite{apruzzese2022modeling}. 

Recent research has focused on designing adversarial network traffic in a black-box setting where the attacker has limited or no information about the defender's NIDS \cite{apruzzese2020deep, zhang2022adversarial}. Black-box attacks can be launched using two approaches: model querying \cite{chen2017zoo} and transferability \cite{DBLP:journals/corr/GoodfellowSS14}.

Studies have shown that by sending multiple queries to the target NIDS, an attacker can extract useful information such as the classifier's decision, which allows the attacker to craft adversarial perturbations capable of evading the NIDS \cite{apruzzese2020deep,zhang2019evasion,ren2020query}. Although achieving high success rates, this approach has two limitations: the first is that NIDSs are not designed to give feedback when queried, unless side-channel attacks are used \cite{boenisch2021side}. The second limitation is that it requires a relatively high number of queries to work properly, which can expose the attacker to be easily detected using a simple query detector \cite{zhang2020tiki, zhang2022adversarial}. Thus, the IDS querying approach is less feasible for a realistic attack \cite{apruzzese2022modeling}.

The second black-box approach is to rely on the transferability property of adversarial examples. This property was first explored by Goodfellow et al \cite{DBLP:journals/corr/GoodfellowSS14}, who show that adversarial examples that fool one model can fool other models with a high probability without the need to have the same architecture or be trained on the same dataset. An attacker can exploit the transferability property to launch black-box attacks by creating a surrogate model trained on data following the same distribution as the target model \cite{debicha2021detect}. This can be easily done by sniffing network traffic \cite{al2020real}, especially in a botnet scenario where the attacker already has a foothold in the corporate network. 

Miller et al. \cite{miller2020adversarial} classify defenses against evasion attacks into two groups: robust classification and anomaly detection. The goal of robust classification is to accurately classify all adversarial instances (i.e., classify them as malicious in our case) while trying not to affect the performance of the classifier in the absence of adversarial attacks. The anomaly detection strategy, on the other hand, advocates detecting and treating adversarial instances rather than attempting to accurately classify them despite the attack.

Adversarial training \cite{DBLP:journals/corr/SzegedyZSBEGF13,DBLP:journals/corr/GoodfellowSS14} was one of the first robust classification strategies to be proposed. Adversarial examples are added to the training set used to create the classifier as part of the data augmentation technique known as adversarial training. The correct label is given to these adversarial data in order to build a more robust classifier that can correctly classify samples even if they have been adversarially modified.
This method, while showing good prospects in some areas of research, is subject to the problem of blind spots \cite{miller2020adversarial}, which means that if adversarial examples of a certain attack family are not included in the learning phase, the classifier may fail to be robust to these adversarial examples. Consequently, this implies that adversarial training could not be as effective against zero-day attacks, which are one of the key issues facing the intrusion detection sector. Moreover, the IDS may perform worse in its original classification job as a result of being retrained with adversarial instances \cite{DBLP:journals/corr/abs-2104-09852}.

Defensive distillation, which was originally used to lower the dimensionality of DNNs, is another adversarial defense presented in the literature \cite{DBLP:conf/sp/PapernotM0JS16, DBLP:journals/tetci/ApruzzeseACM20}. Papernot et al. \cite{DBLP:conf/sp/PapernotM0JS16} suggest a distillation-based defense with the goal of smoothing the model's decision surface. Carlini and Wagner \cite{carlini2016defensive} demonstrated that defensive distillation is no more robust to adversity than unprotected neural networks. As such, this defense is broken using the C\&W attack.

To defend against adversarial assaults, one might discard any features that could be modified by the attacker without affecting the logic of network communication \cite{DBLP:journals/tcyb/ZhangCBYR16, DBLP:conf/acsac/SmutzS12}. However, Apruzzese et al. \cite{apruzzese2019evaluating} demonstrated that doing so has a negative influence on detector performance in non-adversarial conditions, for example, by generating more false alarms, which is extremely undesired for current cyber security platforms. This decrease in performance is due to the deleted features having a major influence on the underlying processes of the baseline detectors.

Anomaly detection is an interesting alternative defensive mechanism against evasion attacks. One of the noted advantages is that correctly classifying adversarial examples is difficult, while detection is slightly simpler \cite{DBLP:conf/ccs/Carlini017}. A supporting argument is that if a good robust classifier is created, then a good detector is obtained for free: detection can be performed when the decision of the robust classifier differs from the decision of the non-robust classifier \cite{miller2020adversarial}. Instead of trying to correctly classify the adversarial instance into its original class, this strategy recommends detecting the adversarial instance and treating it accordingly. The most common action is to reject the identified adversarial sample and prevent it from passing through the system \cite{DBLP:conf/iccv/LuIF17,DBLP:journals/neco/MillerWK19, debicha2021detect}. Our proposed defense falls into the category of anomaly detection.

\section{Proposed method}
\label{sec:proposed}
\subsection{Threat  scenario}
\label{subsec:threatsceanario}
With the increasing use of machine learning-based NIDS, some research has focused on the possibility of evading these NIDS using adversarial attacks already well known in the field of computer vision, which have shown various weaknesses inherent in machine learning algorithms. In order to overcome these weaknesses, it is therefore essential to focus on the robustness of machine learning techniques in the cybersecurity domain.

As shown in Figure \ref{fig:botnetillust}, our work is based on a realistic scenario involving a connected computer device in an enterprise network infected with malware, making it part of the botnet controlled by a dedicated centralized server called Command and Control (C\&C), which is designed to command a multitude of bots. Within this network, a NIDS using a neural network-based model is present to detect any form of attack on the network. In high-speed networks environment and considering the difficulties of analyzing each individual packet, it is realistic to consider that the NIDS is a flow-based system rather than a packet-based system. This NIDS therefore analyzes the flow data generated by the router based on the traffic outgoing/entering the network. All flows first pass through a flow exporter, which extracts the network features, as described in Table \ref{tab:feature_descr}, containing the information of each network flow and sends it to the NIDS for preprocessing and classification. In this table, three groups can be highlighted:

\begin{itemize}
    \item The first one, with a green color, contains all the features that can be directly manipulated by an attacker.

    \item The second one, highlighted in yellow, contains the features depending on the first group, and can therefore be manipulated indirectly.

    \item The last group, with a red background, contains features that cannot be manipulated by an attacker.
\end{itemize}

\begin{figure}[ht]
\centering
\includegraphics[width=\columnwidth]{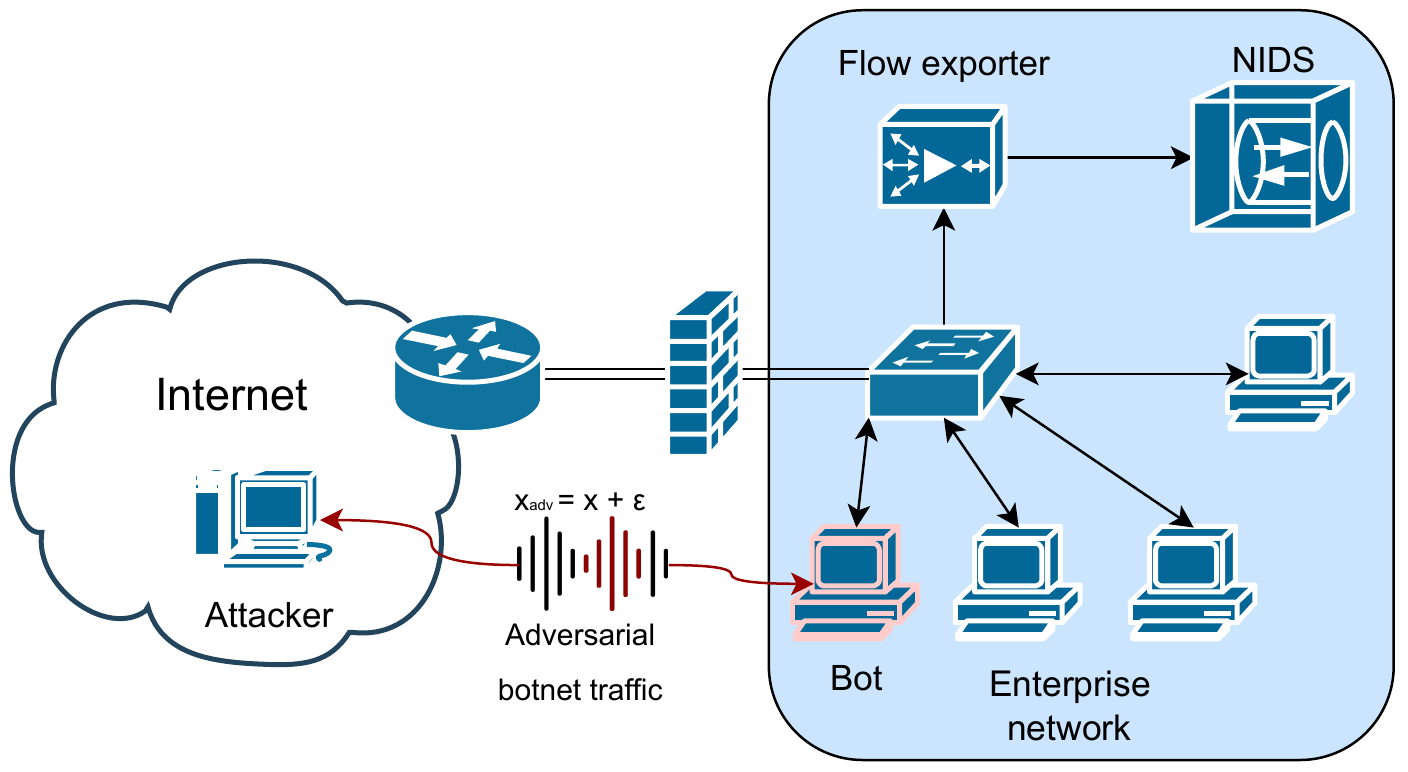}
\caption{An illustration of the considered threat scenario}
\label{fig:botnetillust}
\end{figure}

\begin{table}[ht]
    \centering
    \caption{Common features description used by flow-based NIDS for botnet attacks \cite{venturi2021drelab}}
    \includegraphics[width=\columnwidth]{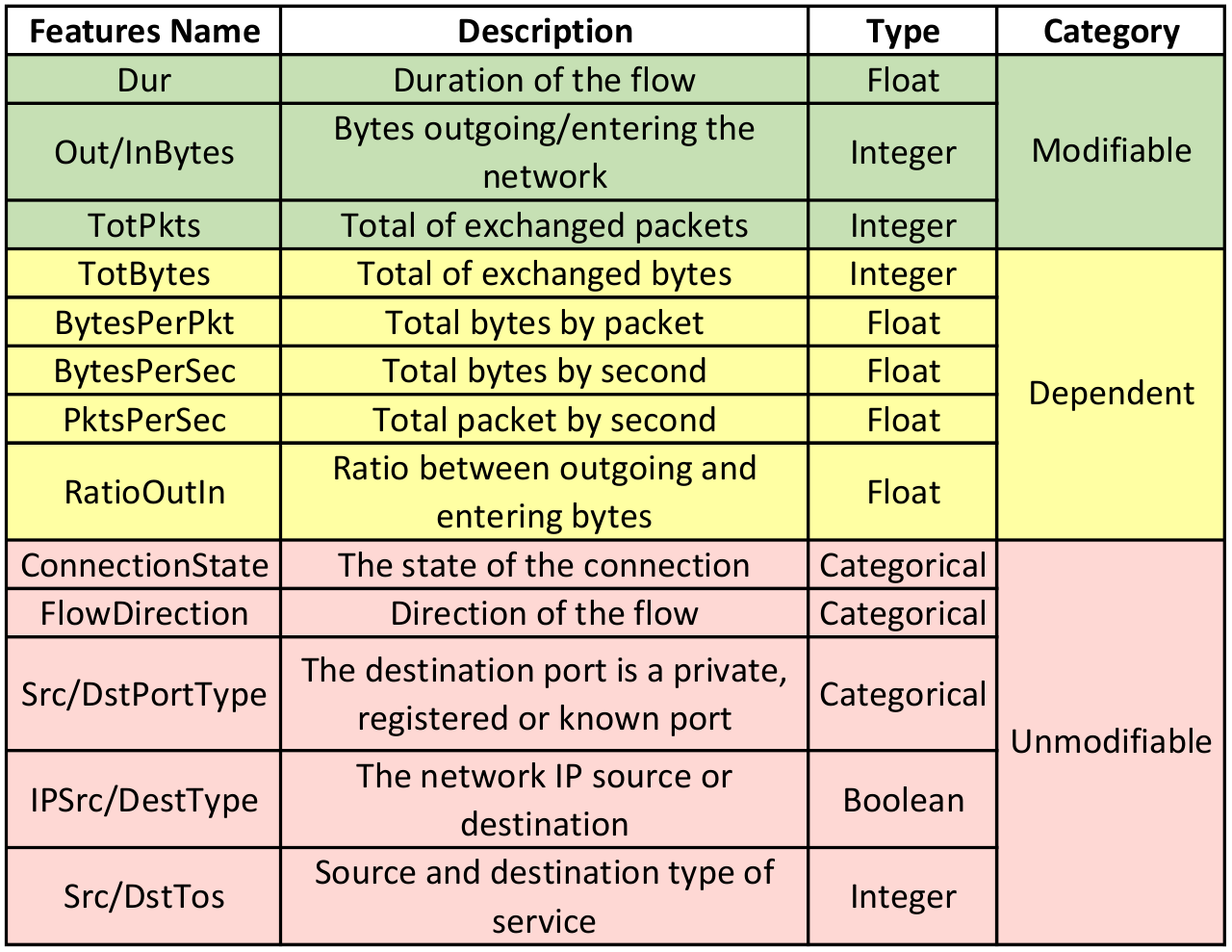}
    \label{tab:feature_descr}
\end{table}

Concerning the temporal aspect of the execution, the processing of the network flow is done in real time on the NIDS, which processes the information online while its learning phase has been carried out offline.

Concerning the knowledge assumptions, it is considered that the attacker has limited access to network data and has no information about the model and parameters of the NIDS used by the company. This means he can only intercept traffic that passes through his machine and gathers limited information about the benign traffic that passes through the bot's machine. In terms of requirements, our scenario assumes that the attacker has previously breached the network by infecting a machine with malware in order to connect to the C\&C server. As a result, the attacker will be able to gather information about the benign traffic. Regarding the flow exporter, the attacker does not need to have knowledge about it or about the extracted features used by the NIDS since he acts directly on factors that he can manipulate in the traffic space. Given the recurring use of certain network factors (in particular, packet duration, number, and size) in botnet detection by state-of-the-art NIDSs, the attacker can assume that these factors are part of the set of features used by the NIDS. The attacker can retrieve these feature lists from known exporters such as Argus \footnote{https://openargus.org/}, CIC-FlowMeter \footnote{https://github.com/CanadianInstituteForCybersecurity/CICFlowMeter} or nProbe \footnote{https://www.ntop.org/products/netflow/nprobe/} and from scientific papers \cite{sarhan2022towards, pektacs2017effective,apruzzese2019evaluating} explaining the features used by NIDS models. The attacker has the ability to communicate with and target the infected computing device. It is therefore also considered that the attacker can manipulate, in both directions of communication, \textit{the duration of packet transmission}, \textit{the number of packets sent and received}, as well as\textit{ the number of bytes sent and received}, as shown in Table \ref{tab:feature_descr}, respecting the semantic and syntactic constraints related to the network protocols used and maintaining the underlying logic of his malicious communications. In order to maintain the malware's full behavior, the attacker cannot directly act on certain features, such as the source and destination IP addresses or the type of service. In fact, the features highlighted in red in Table \ref{tab:feature_descr} cannot be modified by the attacker, either directly or indirectly. Only the three green features: Dur, Out/InBytes, and TotPkts can be manipulated by the attacker. It should be noted that modifying the green features will result in some indirect changes to the yellow features. These changes should be taken into account in order to properly address the respect of semantic and syntactic constraints. To manipulate the network factors explained just before, the attacker can use the following three approaches:

\begin{enumerate}
    \item Time manipulation attack: with this approach, it is possible to act on two time-related aspects during the attack. On the one hand, by reducing the frequency of the attack packets by increasing the sending time between packets of the same flow, as in the work of Zhijun et al. \cite{zhijun2020low} which shows the implication of this variant on DoS attacks. On the other hand, by accelerating the frequency of attack packets in a moderate way by decreasing the time taken to send the packets. These two variants allow to influence directly the "Duration" feature and indirectly the "BytesPerSec" and "PktsPerSec" features.
    
    \item Packet quantity manipulation attack: This attack can be carried out in two different manners. The first is packet injection, which suggests injecting new packets into the network flow by creating them directly, with tools such as Hping \footnote{http://www.hping.org/} or Scapy \footnote{https://scapy.net/}, or by breaking a packet into several fragments, using packet fragmentation, so as to preserve the content and underlying logic of the packet without damaging its overall behavior. Packet fragmentation is, for example, used by some attacks such as the TCP fragment attack \cite{ziemba1995rfc1858} or a DoS attack. A variant suggested by this possibility would be to resend the same packet multiple times using a tool like Tcpreplay \footnote{https://tcpreplay.appneta.com/}. The other way to do this is packet retention, which would be to not send a packet immediately, but rather to send it after a certain amount of time, thus adding that packet to the next flow, thus avoiding having more packets in a single flow. With the same idea of retention, another suggested possibility could be a modification of the general communication system used in the botnet attack to shorten the number of packets sent between the botnet devices. In both cases, this attack directly influences the "TotPkts" feature, and indirectly the "PktsPerSec" and "BytesPerPkt" features.
    
    \item Byte quantity manipulation attack: Similar to the previous approach, there are two ways to perform this attack. The first is byte expansion. In the case where the communication is not encrypted, a suggestion would be to directly modify the payload to obtain the desired number of bytes. In case it is encrypted, it is assumed that the attacker knows the cryptographic suite used to encrypt the two-way communication channel. This would allow him to add padding using a tool like Scapy, calculate its encrypted version, and then verify that the packet size is the desired one. The addition of padding is known to both communicating parties, allowing the receiver to remove the unnecessary part and thus recover the original payload. The second method is byte squeezing. The idea is to reduce the number of bytes sent. To do this, it would be possible to compress the data to directly reduce the size of the payload using a tool like Zlib \cite{gailly2012zlib}. This is a common tool already used in the IoT domain to reduce the bytes exchanged, as explained in \cite{zahra2020packet}. Another possibility would be to modify the general behavior of the botnet system by minimizing the content of the payload to be sent. This second possibility can only be done before the machine is affected by the malicious bot software, but can be proactively prepared by an attacker. Furthermore, whether it is byte expansion or byte compression, the attack directly influences the "OutBytes" and "InBytes" features, and indirectly the TotBytes, "BytesPerPkt", "BytesPerSec" and "RatioOutIn" features.

\end{enumerate}

\begin{figure}[ht]
\centering
\includegraphics[width=\columnwidth]{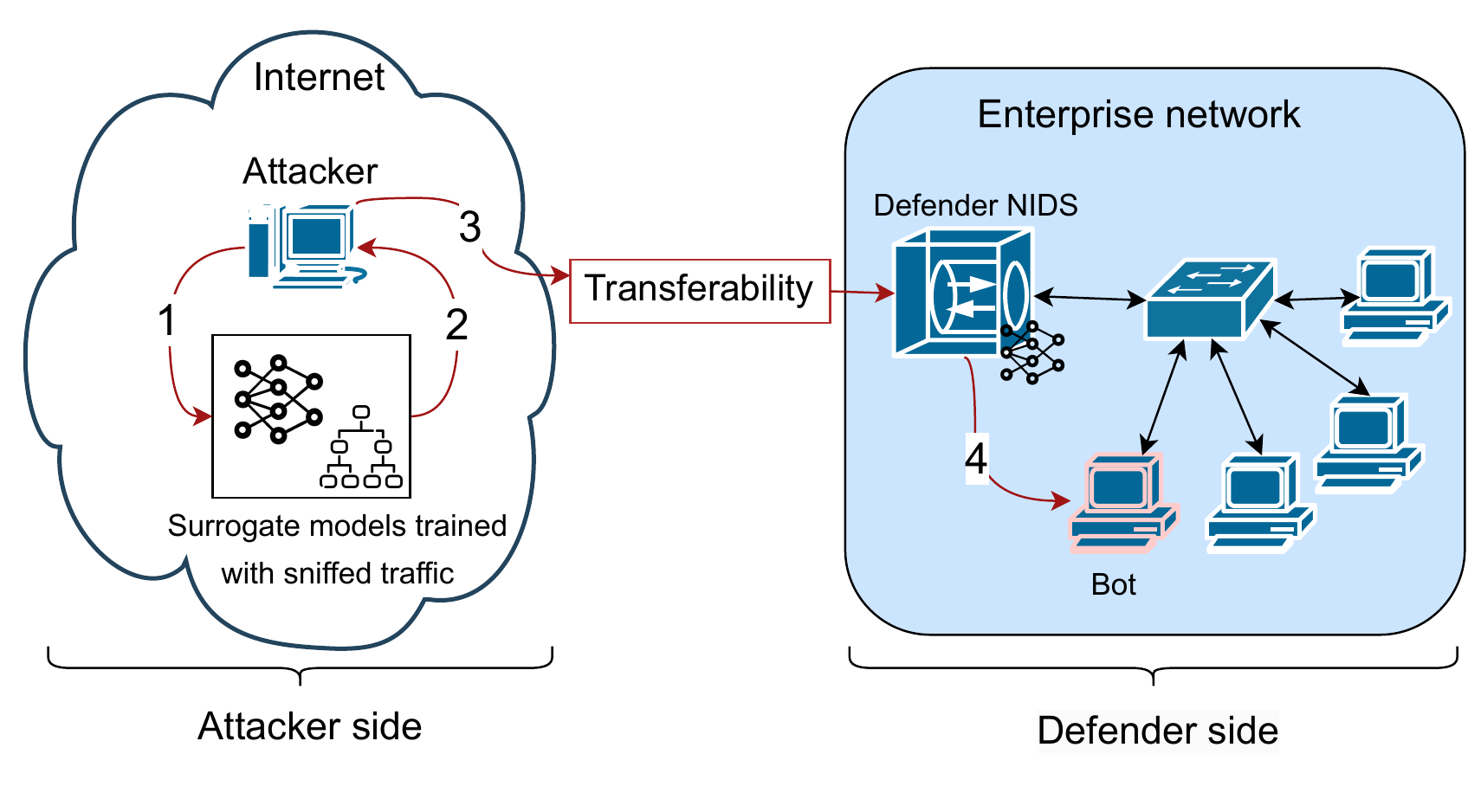}
\caption{An illustration of the adversarial botnet traffic generation steps}
\label{fig:botnetillust}
\end{figure}

 As shown in Figure \ref{fig:botnetillust}, There are four steps in the process of creating adversarial botnet traffic. During step 1, the attacker generates adversarial traffic that is specifically designed to bypass the surrogate models that the attacker previously trained using sniffed traffic. The attacker then receives and analyzes the adversarial traffic that managed to avoid detection by the surrogate models during step 2. During step 3, the attacker uses the transferability property to send adversarial botnet traffic to the defender NIDS. In step 4, the adversarial botnet traffic that successfully bypassed the defender NIDS will arrive at its final destination, the bot.

\subsection{Datasets}

To perform reproducible experiments, the CTU-13 \cite{garcia2014empirical} and CSE-CIC-IDS2018 \cite{DBLP:conf/icissp/SharafaldinLG18} datasets were used to provide results that could be comparable to multiple datasets as well as a wider range of attacks.

\begin{itemize}
    \item CTU-13: This dataset, provided by the Czech Technical University in Prague, contains different labeled traffic from 13 different application scenarios. Each of them follows the same process, including a different botnet attack variant (Neris, Rbot, Virut, Murlo, Menti, NSIS.ay and Sogou). The process involves monitoring a network for both benign network communications and malicious traffic executed by the botnet attack. All of this traffic is extracted as PCAP files and then formatted as flows via the Argus tool, which turns raw network data into flows.
    
    \item CSE-CIC-IDS2018: This dataset contains a set of computer attack scenarios, such as Brute-force, Heartbleed, Botnet, DoS, DDoS, or Web attacks. There are two variants of botnet attacks: Zeus and Ares. The network traffic for each scenario was extracted as a PCAP file and formatted by CICFlowMeter to provide a set of 83 network features for thousands of labeled network flows. This dataset, being relatively recent, provides consistent data following the dataset creation methodology present in \cite{gharib2016evaluation}, allowing to have a reliable and realistic dataset. 
    
\end{itemize}

Other than botnets, CTU-13 and CSE-CIC-IDS2018 contain a variety of attack types. Because our research focuses solely on botnet attacks, all other attack types were removed to create a dataset containing only botnet attack records. To avoid incorporating potential biases when creating these new datasets, we relied on the work done by Venturi et al. \cite{venturi2021drelab}. Features were filtered to keep only those consistent for the study of botnet attacks and common to both datasets used. These features are those described in Table \ref{tab:feature_descr}. Regarding CTU-13, the botnet attacks considered in this work are Neris, Virut, and Rbot. Other attacks were not included because they do not have enough malicious instances to provide consistent results. For CSE-CIC-IDS2018, after initially being indistinguishable in the original dataset, the Zeus and Ares botnet attacks were extracted into the same dataset (Zeus \& Ares).

\begin{figure}[ht]
\centering
\includegraphics[width=\columnwidth]{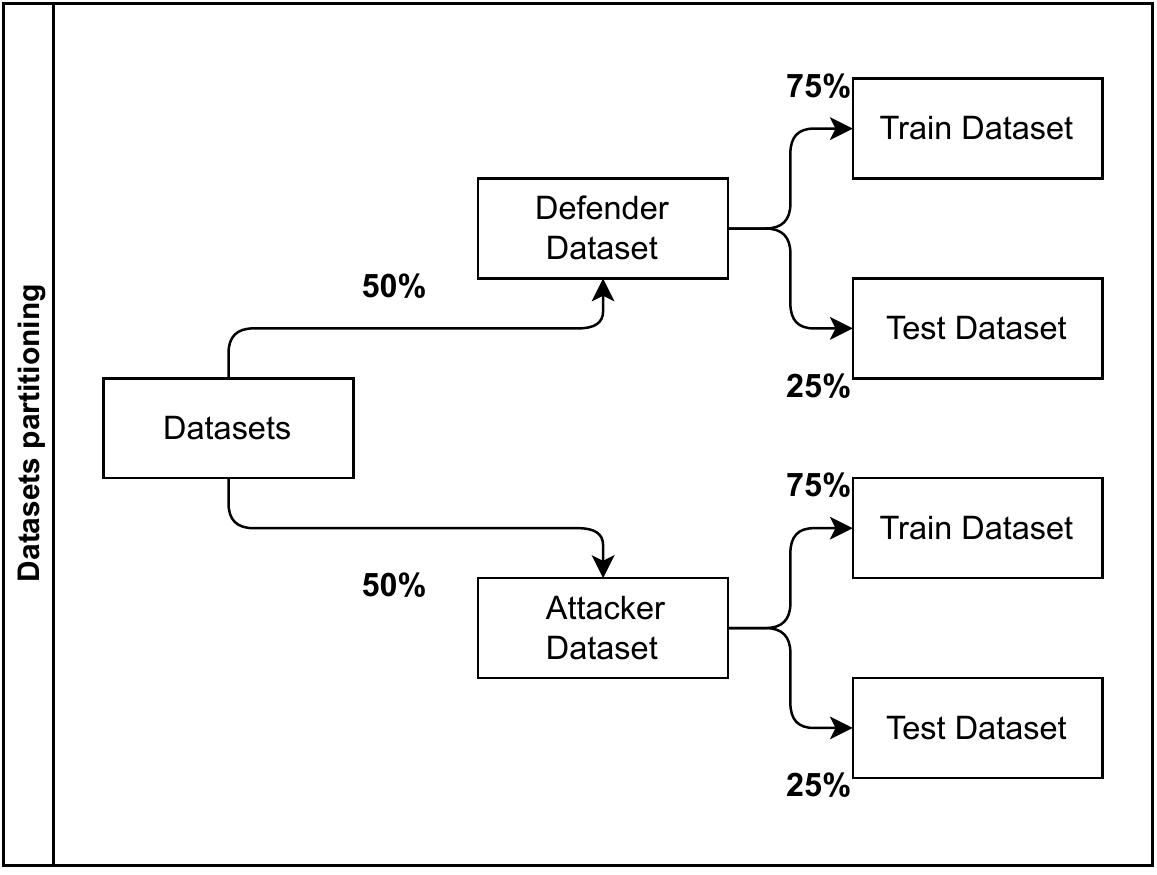}
\caption{Partitioning of the datasets for experiments}
\label{fig:dataset_disposition}
\end{figure}

To ensure the practicality of the present work, the CTU-13 and CSE-CIC-IDS2018 datasets were divided into two equivalent datasets and stratified according to the labels. These datasets are equivalent in terms of size and distribution, which represents more than 32,000 instances for each side. The first is used for training and evaluation of the model used by the defender. The second is used by the attacker to train the surrogate models independently. The attacker can obtain this data by sniffing the network. This is particularly possible in the case of a botnet scenario, as the attacker communicates bidirectionally with the infected device. The datasets for each side (defender and attacker) are split into a training dataset and a test dataset for validation with proportions of 75\% and 25\% respectively. Each training and test data subset is evenly split in terms of malicious and benign traffic. The datasets are separated in this manner to have the most balanced representation so as to avoid the problem of unbalanced data given that it is out of the scope of this study. The attacker and defender datasets thus follow similar but not identical distributions since they are not the same datasets. The arrangement of the datasets is illustrated in Figure \ref{fig:dataset_disposition}.

\subsection{Preprocessing}

General preprocessing has already been performed on the dataset provided by Venturi et al. \cite{venturi2021drelab}, resulting in clean data where outliers, empty values, and non-TCP traffic were removed. Data filtering and "one-hot encoding" of categorical features were carried out. This encoding transforms the categorical data into a binary format that can be exploited by the model. However, some inconsistencies are present in the dataset provided by Venturi et al. \cite{venturi2021drelab}. By default, when the "OutBytes" and "InBytes" features are set to 0, the "RatioOutIn" feature is set to the maximum value present in the data set, simulating an infinite value. Since the ratio in this case is 0/0, we chose to replace it with 0 instead of a pseudo-infinite value, representing a zero byte exchange.

For training the neural network algorithms, some additional preprocessing was applied to the training and test data. First, the data were normalized using a minmax scaling method, transforming the data to be projected to a value between zero and one, according to Eq. \ref{eq:minmaxscale}. Then, the labels undergo a one-hot encoding to transform them to binary values so that they can be processed by the MLP.

\begin{equation}
\label{eq:minmaxscale}
    X\_scaled = \frac{X - min(X)}{max(X) - min(X)}
\end{equation}

where $X$ is an instance and $X\_scaled$ is the normalized instance. 

\subsection{ML-based NIDS}

\begin{figure}[ht]
\centering
\includegraphics[width=\columnwidth]{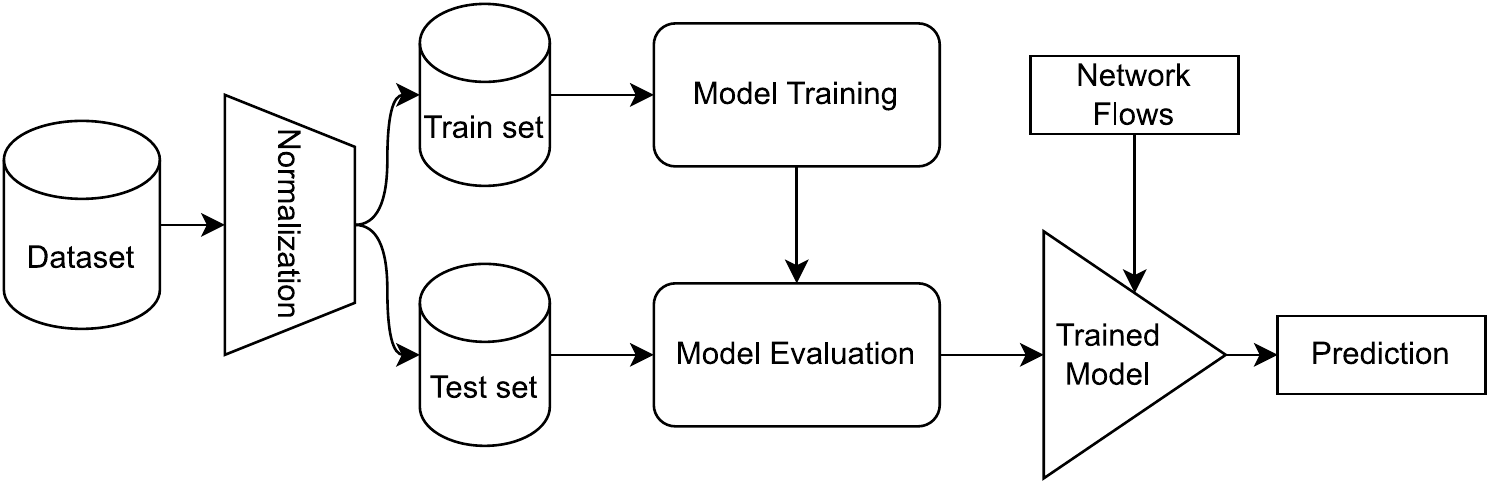}
\caption{Machine Learning pipeline for both attacker and defender}
\label{fig:mlpipeline}
\end{figure}

As neural networks are increasingly used in the context of NIDS based on machine learning algorithms, a couple of them are chosen, as well as more classical ML algorithms. On the defender side, the defender uses a Multilayer Perceptron (MLP), Random Forrest (RF) and K-Nearest Neighbors (KNN) algorithm as a model for his IDS. For the attacker, the same algorithms are chosen with different parameters and hyperparameters and trained with a different dataset. These three ML models were chosen in our work given their popular use in the IDS community \cite{wang2020dynamic,min2018tr,wazirali2020improved}. All these algorithms follow the same training and testing process, as shown in Figure \ref{fig:mlpipeline}.

Meta-parameters are chosen randomly to avoid having the same model parameters between the attacker and defender. Examples of meta-parameters are the the number of neighbors k defined in the KNN algorithm, the number of hidden layers or neurons in the DNN, or the number of estimators used in the RF. The choice to use several algorithms allows for more comparable results, especially in the case of transferability. Having several ML models allows for a better understanding of the impact of transferability of adversarial network traffic, both intra and cross-transferability. The meta-parameters of all models can be seen in Table \ref{tab:modelparam}. 

\begin{table}[ht]
\caption{Meta-parameters of the selected ML models}
\begin{adjustbox}{width=\columnwidth,center} 
\begin{tabular}{|c|c|c|}
\hline
\textbf{Classifier} &
  \textbf{Attacker Parameters} &
  \textbf{Defender Parameters} \\ \hline
MLP &
  \begin{tabular}[c]{@{}c@{}} \textbf{Hidden Layers (HL) = 3}\\ \textbf{Neurons by HL = 128}\\ Activation = ReLU\\ Optimizer = Adam\end{tabular} &
  \begin{tabular}[c]{@{}c@{}} \textbf{Hidden Layers (HL) = 2}\\ \textbf{Neurons by HL = 256}\\ Activation = ReLU\\  Optimizer = Adam\end{tabular} \\ \hline
Random Forest &
  \begin{tabular}[c]{@{}c@{}}\textbf{Nb estimators = 300}\\  Criterion = Gini\\ Bootstrap = True\end{tabular} &
  \begin{tabular}[c]{@{}c@{}}\textbf{Nb estimators = 200}\\  Criterion = Gini\\  Bootstrap = True\end{tabular} \\ \hline
KNN &
  \textbf{Nb neighbors = 5} &
  \textbf{Nb neighbors = 3} \\ \hline
\end{tabular}
\end{adjustbox}
\label{tab:modelparam}
\end{table}

\subsection{Proposed evasion attacks }

In order to generate adversarial perturbations, which are added to the malicious instances to make them benign, we propose two evasion attack variations formulated in Eq. \ref{eq:attackmeanratio} and Eq. \ref{eq:attackmeandiff} .

\begin{equation}
\label{eq:attackmeanratio}
\begin{aligned}
    x^t_{adv}(f) = Proj[x^{t-1}(f) + sign(benign\_mean(f) \\ -  x^0(f)) * (c*t) * mean\_ratio(f)]
\end{aligned}
\end{equation}
where $x^0(f)$ is the initial value of the targeted network factor $f$ in the instance, the sign function specifies the direction of the perturbation, $c$ is a multiplicative coefficient that regulates the perturbation rate generated at each step $t$, $benign\_mean$ is the mean of the benign set of the targeted network factor f that the attacker can obtain from sniffing network traffic, $mean\_ratio$ is the ratio between the mean of the malicious set and the benign set of the targeted network factor $f$, and $Proj$ is a projection function that projects the values of modified features that violate syntactic and semantic constraints into the space of valid values.  These modified features are only network factors that the attacker can manipulate directly or indirectly, as represented by the green and yellow groups in Table \ref{tab:feature_descr}. If the attacker manipulates one of the modifiable features shown in Table \ref{tab:feature_descr} with green color, the \textit{Proj} function will make sure that the dependents features will change values accordingly. For instance, changing the flow duration by the attacker will induce a change in "total packet by second" feature which equals number of packets divided by the flow duration. In nutshul, the projection function is what allows, when features are modified, to ensure that the domain constraints are respected. This enables the adversarial instances created to be valid and reversible in the targeted domain. In this manner, the malware's intended behavior is preserved.

\begin{equation}
\label{eq:attackmeandiff}
\begin{aligned}
    x^t_{adv}(f) = Proj[x^{t-1}(f) + sign(benign\_mean(f)\\ - x^0(f)) * (c*t) * |mean\_diff(f)|]
\end{aligned}
\end{equation}
where $mean\_diff$ is the difference between the mean of the benign set and the malicious set of the targeted network factor $f$. In this case, $mean\_diff$ is an absolute value to avoid influencing the direction of added perturbations during adversarial generation.

\begin{figure}[ht]
\centering
\includegraphics[width=0.85\columnwidth]{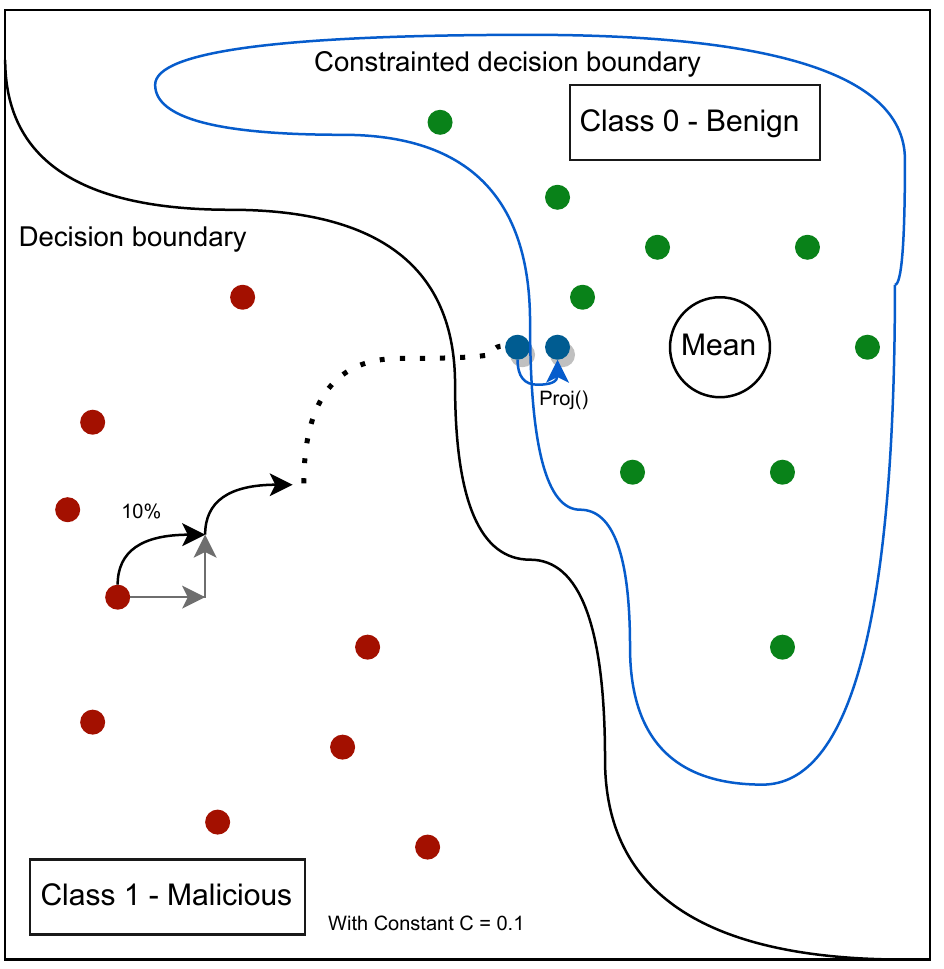}
\caption{2D illustration of a malicious instance transformation with the mean difference method}
\label{fig:advtransfo}
\end{figure}

Although both methods yield similar results, we decided to use the second method (i.e. mean difference method as in Eq. \ref{eq:attackmeandiff}) to conduct the experiments. This choice is purely for illustrative purposes, as the mean difference method is more intuitive and can be illustrated by Figure \ref{fig:advtransfo} which shows how a malicious instance tends to become benign, and thus adversarial. Mean is the average of the benign data. The figure is shown in two dimensions for the sake of simplicity, but in reality the manipulation space is much larger because many factors are being manipulated. 
 
\subsection{Adversarial instances generation}

\begin{figure}[ht]
\centering
\includegraphics[width=\columnwidth]{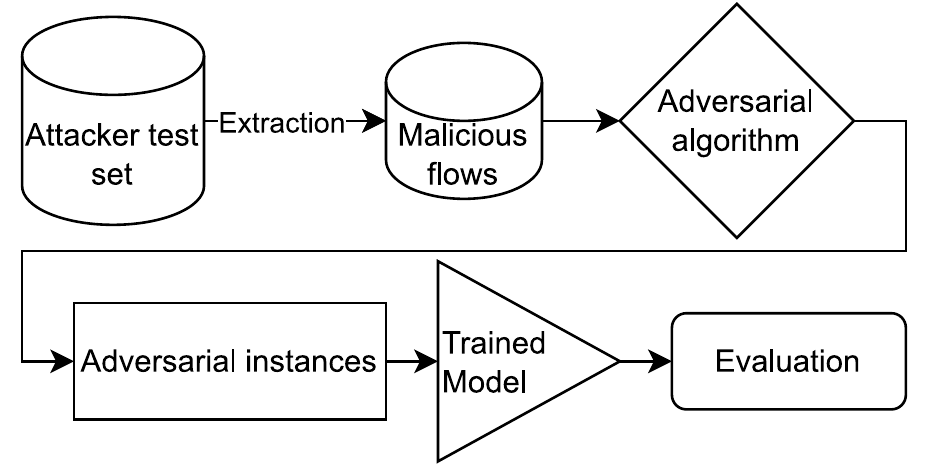}
\caption{Process to generate adversarial instances}
\label{fig:advgeneration}
\end{figure}

To generate adversarial instances, the process illustrated in Figure \ref{fig:advgeneration} is followed. During this process, the previously defined Eq. \ref{eq:attackmeandiff} is applied during the execution of the adversarial generation step described in Algorithm \ref{alg:craftadvexalgo}. 

\begin{algorithm}[hbt!]
\caption{Crafting adversarial examples for flow-based IDS}
\begin{algorithmic}
\Procedure{CraftAdvEx}{x} \Comment{where x is a malicious flow}
\State $x_{adv} \gets x$ 
\State $t \gets 1$ 
\State  $m \gets mean\_difference()$ \Comment{or mean\_ratio()}
\Repeat
    \For{$mask \gets mask_1, ..., mask_{15}$}
        \State $\epsilon \gets sign[benign\_mean(f) - x^0(f)] * (c*t) * m(f)$
        \State $\epsilon \gets \epsilon * mask$
        \State $x_{adv} \gets x_{adv} + \epsilon$ 
        \State $x_{adv} \gets Proj(x_{adv})$
        \If{$predict(x_{adv}) == benign$ }
             \State \textbf{return} $x_{adv}$
        \EndIf
    \EndFor
    \State $t \gets t+1$ 
\Until{$predict(x_{adv}) == benign$}
\EndProcedure
\\
\Procedure{Proj}{$x_{adv}$} 
\Comment{applying  domain constraints to $x_{adv}$}
    \State $x_{adv} \gets ApplySyntacticConstraints(x_{adv})$
    \State $x_{adv} \gets ApplySemanticConstraints(x_{adv})$ 
    \State \textbf{return} $x_{adv}$
\EndProcedure

\end{algorithmic}
\label{alg:craftadvexalgo}
\end{algorithm}

It is worth mentioning that this algorithm applies increasingly large perturbations to the initial malicious stream at each iteration, whenever the new adversarial instance is not classified as benign. Both formulas guarantee small initial perturbations and are modulated by the multiplicative constant $c$.

To select the combinations of attacks (presented in Section \ref{subsec:threatsceanario}) that should be applied to manipulate the network traffic flow, a method of masks is employed where each corresponds to a combination of the manipulable factors (i.e., the total number of packets, the transmission time of the packets, and the number of outgoing or incoming bytes in those packets). At each iteration, the masks are applied in a gradual manner by multiplying them by the previously generated perturbations using Eq. \ref{eq:attackmeandiff} and Eq. \ref{eq:attackmeanratio}. This ensures that only the perturbations needed at that current stage are applied to the selected factors. These masks follow a binary logic and represent 15 combinations. The representation of these masks with their corresponding manipulable factors is shown in Table \ref{tab:combinaisons}. 

\begin{table}[ht]
\caption{Features used by combination with their corresponding mask}
\begin{adjustbox}{width=0.8\columnwidth,center} 
\begin{tabular}{|c|c|c|}
\hline
\textbf{Combinaison} & \textbf{Mask} & \textbf{Target factor}                \\ \hline
1                    & 0001          & Duration                                \\ \hline
2                    & 0010          & TotPackets                              \\ \hline
3                    & 0011          & Duration, TotPackets                    \\ \hline
4                    & 0100          & InBytes                                 \\ \hline
5                    & 0101          & InBytes, Duration                       \\ \hline
6                    & 0110          & InBytes, TotPackets                     \\ \hline
7                    & 0111          & InBytes, TotPackets, Duration           \\ \hline
8                    & 1000          & OutBytes                                \\ \hline
9                    & 1001          & OutBytes, Duration                      \\ \hline
10                   & 1010          & OutBytes, TotPackets                    \\ \hline
11                   & 1011          & OutBytes, TotPackets, Duration          \\ \hline
12                   & 1100          & OutBytes, InBytes                       \\ \hline
13                   & 1101          & OutBytes, InBytes, Duration             \\ \hline
14                   & 1110          & OutBytes, InBytes, TotPackets           \\ \hline
15                   & 1111          & OutBytes, InBytes, TotPackets, Duration \\ \hline
\end{tabular}
\end{adjustbox}
\label{tab:combinaisons}
\end{table}

Once perturbations are applied at each stage, syntactic and semantic constraints are enforced, through the $Proj()$ function, to respect the underlying logic of the malicious network communication. This projection function has a couple of actions: it allows to apply semantic and syntactic constraints to make adversarial instances valid. To do this, this function recalculates the dependent factors (i.e., the yellow group in Table \ref{tab:feature_descr}) that depend on the directly manipulatable factors (i.e., the green group in Table \ref{tab:feature_descr}). Furthermore, when a value obtained exceeds the maximum value listed in the entire test set, this value is projected to this maximum value to ensure that it is transcribed in its appropriate scope. Given that each dependent feature (in yellow) is linked to a modifiable feature (in green) via an easy and intuitive formula explained in Table \ref{tab:feature_descr} (e.g., RationOutIn = OutBytes/InBytes), We chose not to list all of the formulae in Algorithm \ref{alg:craftadvexalgo}, instead providing a list of the corresponding dependent features as shown in Table \ref{tab:semantic_features}, which are updated when an attacker manipulates one of the modifiable features to ensure semantic constraints are respected. This algorithm is feasible in that the attacker acts directly on the network traffic and not on the features. This means that it follows a so-called traffic-based methodology, which is more realistic than the feature-based one because the attacker only needs limited information about the defender, such as information about the network traffic he plans to attack. This allows us to assume black-box knowledge, which is not the case with feature-based manipulation, which requires white-box knowledge, such as access to the internal architecture of the intrusion detection system, its parameters, and the dataset used to train the ML model. It can be noted that our adversarial algorithm was designed to consider two objective functions: one to minimize the amount of modification in each factor and the other to minimize the number of modified factors. 

\begin{table}[ht]
    \centering
    \caption{ When an attacker manipulates one of the modifiable features, the corresponding dependent features are updated using the Proj() function to ensure semantic constraints are respected.}
    \includegraphics[width=\columnwidth]{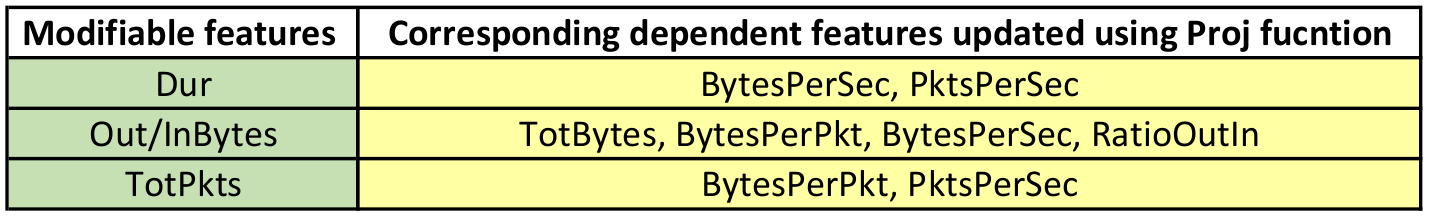}
    \label{tab:semantic_features}
\end{table}

\subsection{Strengthening adversarial robustness}

Since ML algorithms are inherently vulnerable to adversarial examples, it is necessary to find various approaches to defend them. After reviewing a range of defenses found in the literature, we found that many of them are not fully adapted to domain-constrained systems such as NIDS and that there is still room for improvement \cite{carlini2016defensive,xu2017feature}, although some of them have the potential to reduce the impact of adversarial instances \cite{DBLP:journals/corr/GoodfellowSS14,grosse2017statistical}. For this reason, a defense is proposed in this work.

\begin{figure}[ht]
\centering
\includegraphics[width=\columnwidth]{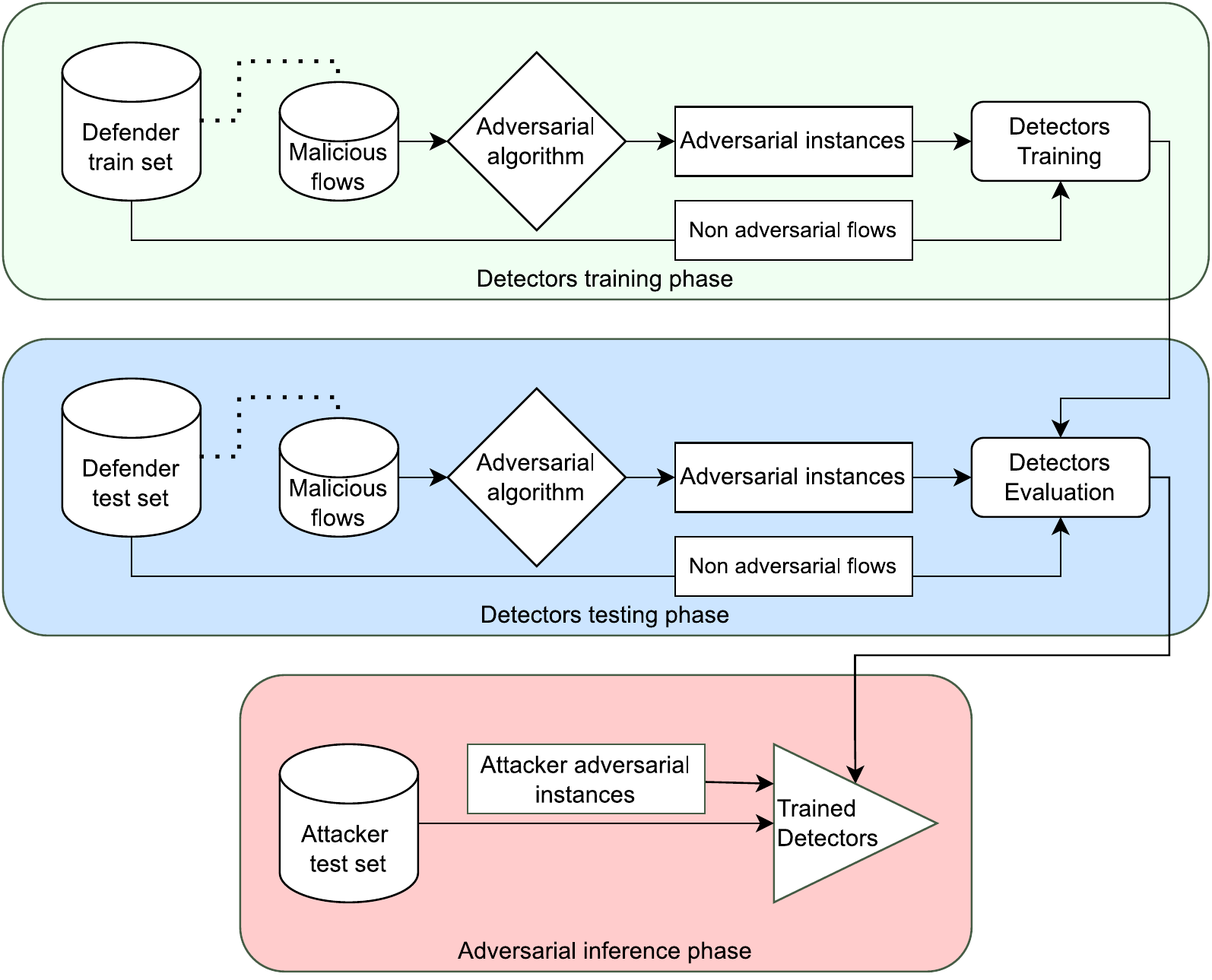}
\caption{Proposed defense process}
\label{fig:defprocess}
\end{figure}

\begin{figure*}[ht]
\centering
\includegraphics[width=0.8\textwidth,height=0.5\textwidth]{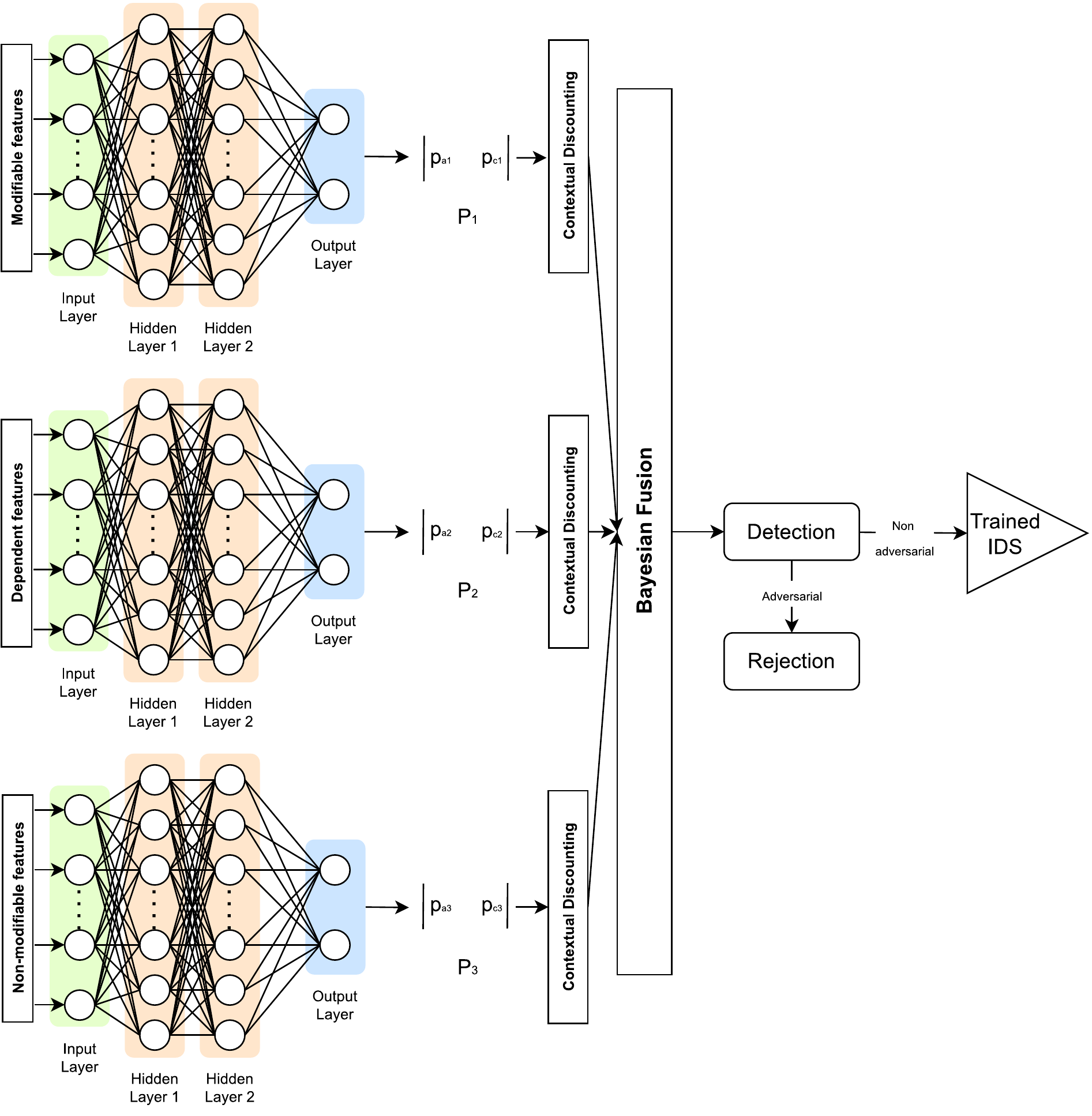}
\caption{Proposed defense approach using MLP sub-detectors}
\label{fig:advdefense}
\end{figure*}
As shown in Figure \ref{fig:advdefense}, the proposed reactive defense follows the process illustrated in Figure \ref{fig:defprocess}. The defense itself is inspired by adversarial training \cite{DBLP:journals/corr/GoodfellowSS14}, adversarial detection \cite{grosse2017statistical}, and bagging as an ensemble method. The role of this defense is to act as an adversarial detector (filter) that is placed before the NIDS to prevent adversarial instances from being able to reach and fool the NIDS. The main idea of this defense is to train three MLP models in parallel, each taking as input a specific group of features in order to detect adversarial instances. There are thus three groups corresponding to features that can be directly manipulated by the attacker, features that depend on the first group, and features that cannot be modified. This arrangement corresponds to the manipulability of features in a realistic context, i.e., features that can be modified by the attacker. The instances used to train the three sub-detectors are divided into two classes: the first one contains benign and malicious instances, having been concatenated together and relabeled as clean (i.e., non-adversarial), the second class contains adversarial instances previously created based on malicious instances exclusively. Table \ref{tab:data_description} illustrates the different data types used to better understand this distinction.

\begin{table}[ht]
\caption{Description of datasets and labels used by the detectors}
\begin{adjustbox}{width=\columnwidth,center}
\begin{tabular}{|c|c|}
\hline
\textbf{Dataset Type} & \textbf{Description}                                    \\ \hline
Benign                & The network traffic that contain benign communications. \\ \hline
Malicious             & The network traffic that contain a botnet attack.       \\ \hline
Clean                 & The concatenation of benign and malicious traffic.      \\ \hline
Adversarial & \begin{tabular}[c]{@{}c@{}}The generated adversarial instances produced \\ with an adversarial algorithm using malicious traffic.\end{tabular} \\ \hline
\end{tabular}
\end{adjustbox}
\label{tab:data_description}
\end{table}

Each of the three feature groups is assigned a particular weight. These weights are set during model training based on the overall detection rate of each detector. Once the predictions have been made by each sub-model for each test instance, they undergo a contextual discounting and then a Bayesian fusion, as defined in Eq. \ref{eq:contdiscount}, where the prediction matrix is multiplied by the corresponding weights, and then the three resulting matrices are summed. Once this step is complete, the new values are normalized to obtain the final probabilities, where $P_a$ represents the probability that the instance is adversarial, and $P_c$ is the probability that it is clean. The result is a final prediction of whether an instance is adversarial or not, and thus is subject to rejection by the detector.

\begin{equation}
\label{eq:contdiscount}
    \begin{split}
P_{a}=\frac{\sum_{i=1}^{3} (P_{a_{i}}*w_{i})}{\sum_{i=1}^{3} (P_{a_{i}}*w_{i})+\sum_{i=1}^{3} (P_{c_{i}}*w_{i})} \\
P_{c}=\frac{\sum_{i=1}^{3} (P_{c_{i}}*w_{i})}{\sum_{i=1}^{3} (P_{a_{i}}*w_{i})+\sum_{i=1}^{3} (P_{c_{i}}*w_{i})}
    \end{split}
\end{equation}

Figure \ref{fig:advdefense} illustrates the use of MLP as a sub-detector, but it should be noted that these detectors can use different machine learning algorithms. This defense could even be used in a stacking scheme involving detectors, each using some specific machine learning algorithms.

\section{Evaluation}
\label{sec:results}

This section discusses the findings of several experiments. First, those concerning the performance of the initial attacker and defender models are discussed based on different metrics, namely precision, recall and F1 score. These results concern the models trained with the CSE-CIC-IDS2018 and CTU-13 datasets. 
In a second step, results regarding the performance of the models in adversarial contexts are discussed. A preliminary study of the models trained with the CSE-CIC-IDS2018 dataset is investigated to see the impact of transferability between the same and different models as well as the training data. A study of the time taken by each attack is also considered, as well as an analysis of the differences in perturbation between the initial malicious instance and the adversarial instance. Then, a general comparison is made on the performance of the proposed adversarial generation algorithm across the botnet attacks present in each dataset. 
The last section includes the results of the proposed defense against the adversarial instances generated by the previously proposed evasion attack algorithm. 

\subsection{Initial performance of ML-IDS models in clean settings}


To evaluate the initial performance of ML-IDS models trained in clean (i.e. non-adversarial) settings on both the attacker and defender sides, several metrics are used, namely: recall defined in Eq. \ref{eq:recall}; precision defined in Eq. \ref{eq:precision}; and f1-score defined in Eq. \ref{eq:f1-score}. ML-IDS models are requested to perform a binary classification to distinguish malicious from benign traffic in clean settings.

\begin{figure}[ht]
    \centering
    \includegraphics[width=\columnwidth]{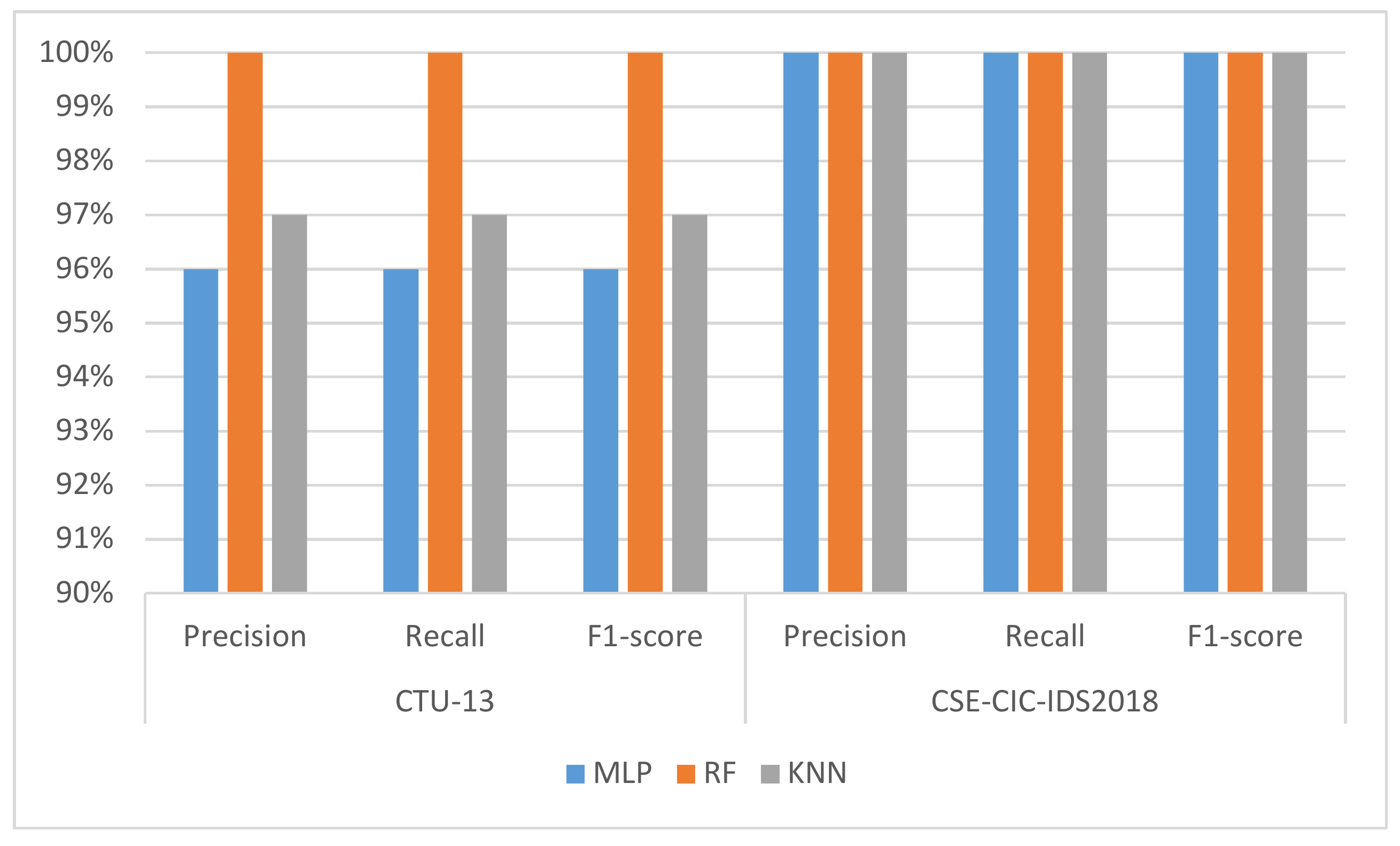}
    \caption{Performance of the attacker-side trained ML models on CTU-13 and CSE-CIC-IDS2018 in clean settings}
    \label{fig:baseline_attacker_models}
\end{figure}

As shown in Figure \ref{fig:baseline_attacker_models}, the results for the initial performance of the attacker-side trained ML models yield metrics of 100\% for all models trained with CSE-CIC-IDS2018. In the case of CTU-13, they are less significant. Nevertheless, all models have metrics above 96\%. These results regarding the performance of ML-IDS on both datasets are comparable to those found in the literature \cite{lima2019smart,kanimozhi2019artificial,nugraha2020performance}. 

These initial results show good performance in general. Models trained with CTU-13 perform somewhat less well than those trained with CSE-CIC-IDS2018. This may be due to the fact that CSE-CIC-IDS2018 contains only two botnet attacks, while CTU-13 contains five, forcing the models to expand their decision boundary to try to correctly classify all types of attacks into a single class (i.e., malicious).

\begin{figure}[ht]
    \centering
    \includegraphics[width=\columnwidth]{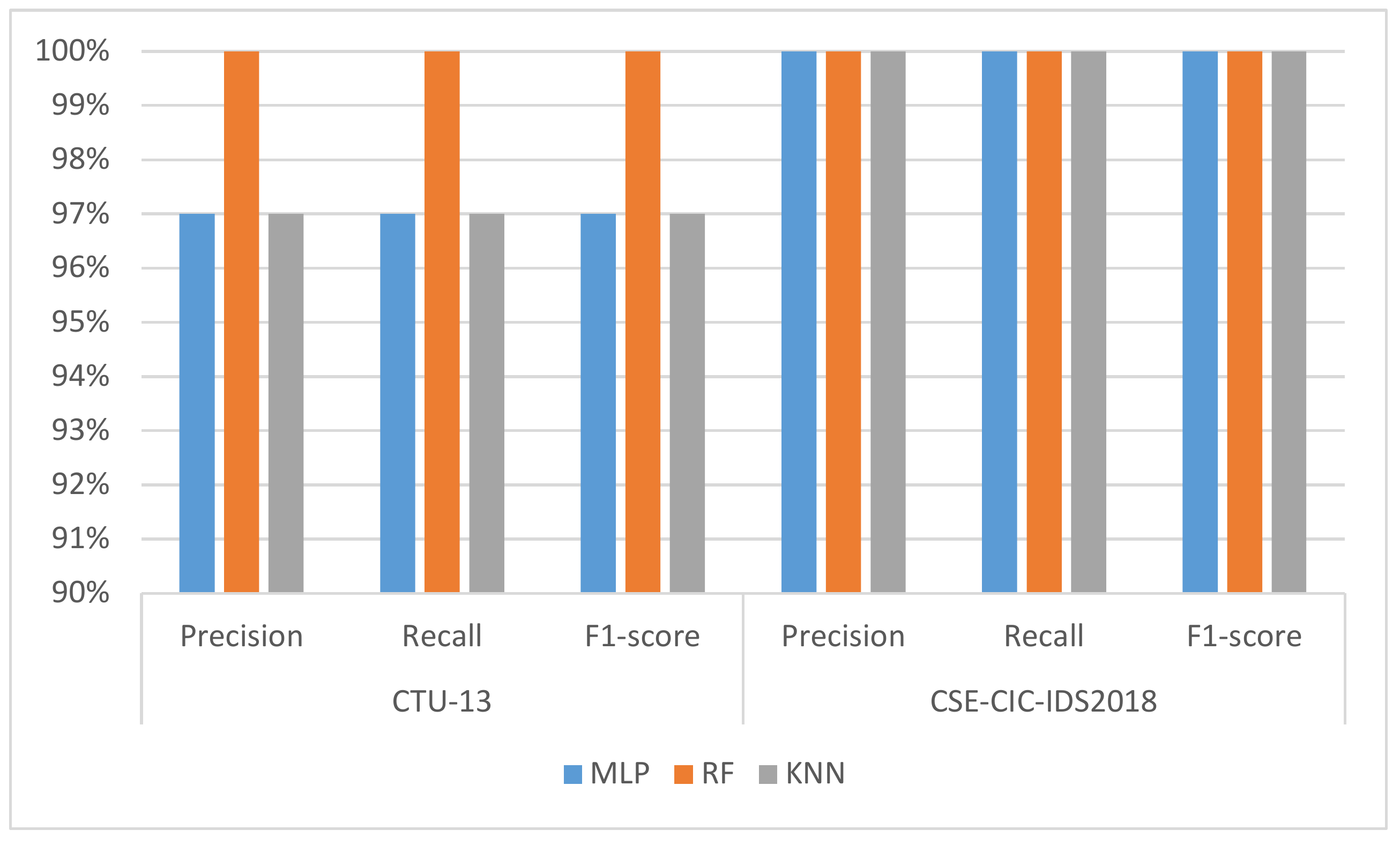}
    \caption{Performance of the defender-side trained ML models on CTU-13 and CSE-CIC-IDS2018 in clean settings}
    \label{fig:baseline_defender_models}
\end{figure}

Regarding the performance of the defender-side trained ML models, presented in Figure \ref{fig:baseline_defender_models}, we can see slight variations in the metrics for the models trained with CTU-13. These variations are very small and of the order of 1\% maximum. We can note that the model using a random forest gives a metric of 100\%. For the models trained with CSE-CIC-IDS2018, the metrics are all at 100\%, which gives identical performance to that of the attacker-side trained ML models. In general, the same observations can be made for the defender models, as the results are relatively similar.


\subsection{Performance of ML-IDS models in adversarial settings}
\label{suc:AdvPerf}

To study the impact of adversarial instances generated by our algorithm \ref{alg:craftadvexalgo}, as well as the effectiveness of transferring these adversarial instances created by the attacker to the models trained by the defender, the first experiment focuses on the CSE-CIC-IDS2018 dataset containing the Zeus \& Ares botnet attacks. To measure the impact of adversarial instances, the detection rate metric, also called recall, is used \ref{eq:recall}. It measures the rate of adversarial instances detected by the ML-IDS as malicious.
To do this, the attacker first generates adversarial instances for each model trained on his side (i.e., MLP, RF, and KNN). The adversarial instances generated for one model will be sent to the other models to evaluate the transferability property between the models trained by the attacker.
The defender-trained models are different from the attacker-trained models in two aspects: i) they have different hyperparameters; ii) they are trained with different datasets.  We test the effect of using the same or different hyperparameters on the transferability of adversarial instances from the attacker's models to the defender's models.

It should be noted that the results of the experiments in the various tables always compare equal proportions for the defender and attacker with different data ( i.e. different contents). As shown in Figure \ref{fig:dataset_disposition}, the size of the training (75\%) and test data (25\%) is the same for the attacker and defender because the original dataset is divided in half. Since the size of the data is the same, by saying that the data is different, we invariably mean its content.

Table \ref{subtab_a:cicidsstudy} illustrates the performance of the adversarial transferability property between the attacker's trained models. It is worth mentioning that all the attacker's models are trained with the same data. As we can see, the diagonal values of the table are all zeros. This is to be expected since the adversarial instances were designed based on the decision boundary of this model, so testing on the same model will give a detection rate of 0\%. We also notice that the adversarial instances generated based on MLP were able to drop the detection rate of RF and KNN to 0\%, while the adversarial instances based on these two models were able to drop the detection rate to about 50\%.  Finally, we can see that on average, the detection rate dropped to 21.9\%, which is a rather satisfactory result for the attacker.

Table \ref{subtab_b:cicidsstudy} represents the impact of adversarial instances, generated using the attacker's model, against the defender's models. The particularity is that the defender uses, in this case, the same hyperparameters as the attacker's models. Only the dataset used during training changes. This allows, compared to Table \ref{subtab_a:cicidsstudy}, to see the effect of intra-transferability through the same models and the same hyperparameters using a different dataset for training, but also of inter-transferability through different models with the same hyperparameters. This gives concrete information about the impact of the training data and the model used on the transferability of adversarial instances. Compared to Table \ref{subtab_a:cicidsstudy}, Table \ref{subtab_b:cicidsstudy} shows slight changes in the effect of intra-transferability (i.e., the same model) as seen in the diagonal values ( 1\% for RF and 3\% for KNN) while the effect of cross-transferability remains similar in both tables. These results indicate that the attacker does not need to train his models with the same data as the defender, data with a similar distribution will be sufficient to make the transferability property work effectively.

Table \ref{subtab_c:cicidsstudy} again shows the adversarial performance against the defender models. However, although the models use the same learning algorithms, this time they have different hyperparameters. This allows us to show the effect of intra-transferability and cross-transferability on different models with different hyperparameters as well as different training data sets. This table is therefore the one that provides the closest insight to the reality because, here, the attacker's knowledge is extremely limited since he neither knows the model used, nor the parameters or hyperparameters of the model, nor the data used to train the defender's model. This demonstrates the effect of transferability in a realistic context. The results provide a fairly similar detection rate between Table \ref{subtab_b:cicidsstudy}  and Table \ref{subtab_c:cicidsstudy}, indicating that not using the same hyperparameters as the defender does not have a significant impact on the transferability of the adversarial instances.  

From Table \ref{tab:cicidsstudy}, we can see that not knowing the training data, hyperparameters, or model used by the defender does not prevent the attacker from creating adversarial instances and successfully transferring them to the defender's ML-IDS.

\begin{table}[ht]
    \caption{Performance of the adversarial transferability between ML-IDS models in term of detection rate}
    \begin{adjustbox}{width=\columnwidth,center} 
    \begin{subtable}{\columnwidth}
        \centering
        \caption{Performance of the adversarial transferability between the attacker's trained models}
        \label{subtab_a:cicidsstudy}
        \includegraphics[width=\columnwidth]{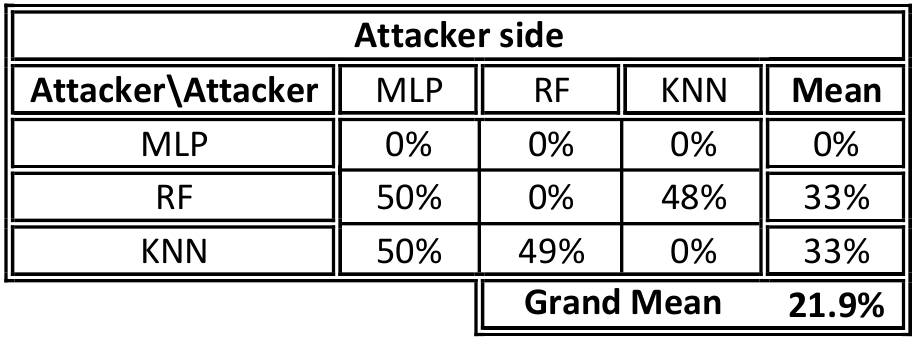}

        \centering
        \caption{Performance of the adversarial transferability from the attacker to the defender models (having the same attacker's hyperparameters)}
        \label{subtab_b:cicidsstudy}
        \includegraphics[width=\columnwidth]{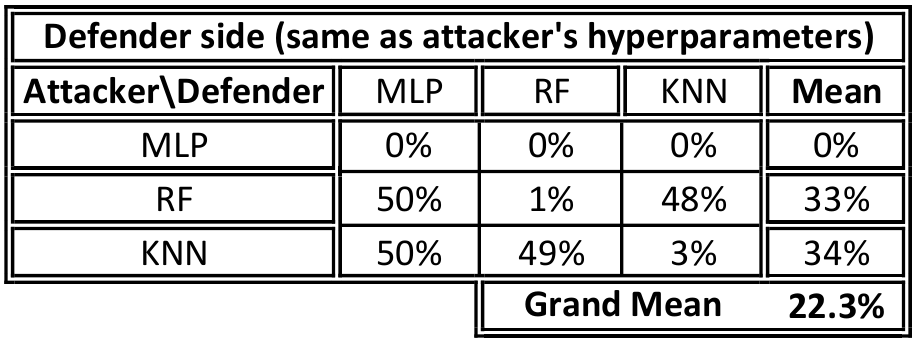}

      \centering
        \caption{Performance of the adversarial transferability from the attacker to the defender models (having different hyperparameters from the attacker's ones)}
        \label{subtab_c:cicidsstudy}
        \includegraphics[width=\columnwidth]{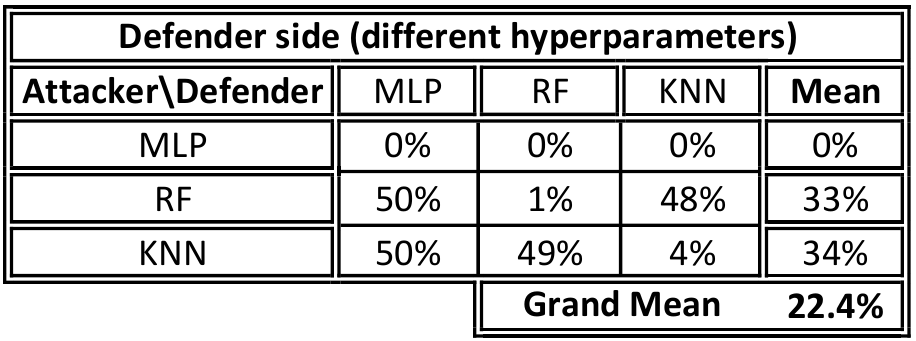}

    \end{subtable}
    \end{adjustbox}
    \label{tab:cicidsstudy}
\end{table}

The set of sub-tables in Table \ref{tab:perturbation_difference} represents the average and maximum perturbation difference between the malicious instances and their adversarial counterparts using the CSE-CIC-IDS2018 dataset. Each of these tables represents the ML model used by the adversarial generation algorithm \ref{alg:craftadvexalgo} to craft the perturbations to be added to the manipulatable factors. Note that the duration factor "Dur" is expressed in seconds.

The main observation that can be made from this set of tables is that the perturbations are rather small for all the manipulable factors and therefore feasible in realistic scenarios. It should be noted that these perturbations are influenced by the number of steps and the regulation coefficient c present in the adversarial generation algorithm \ref{alg:craftadvexalgo} and driven by the mean of each manipulable factor.

\begin{table}[ht]
    \centering
    \caption{The maximum and average perturbation difference between the malicious and adversarial instance on CSE-CIC-IDS2018}
    \includegraphics[width=\columnwidth]{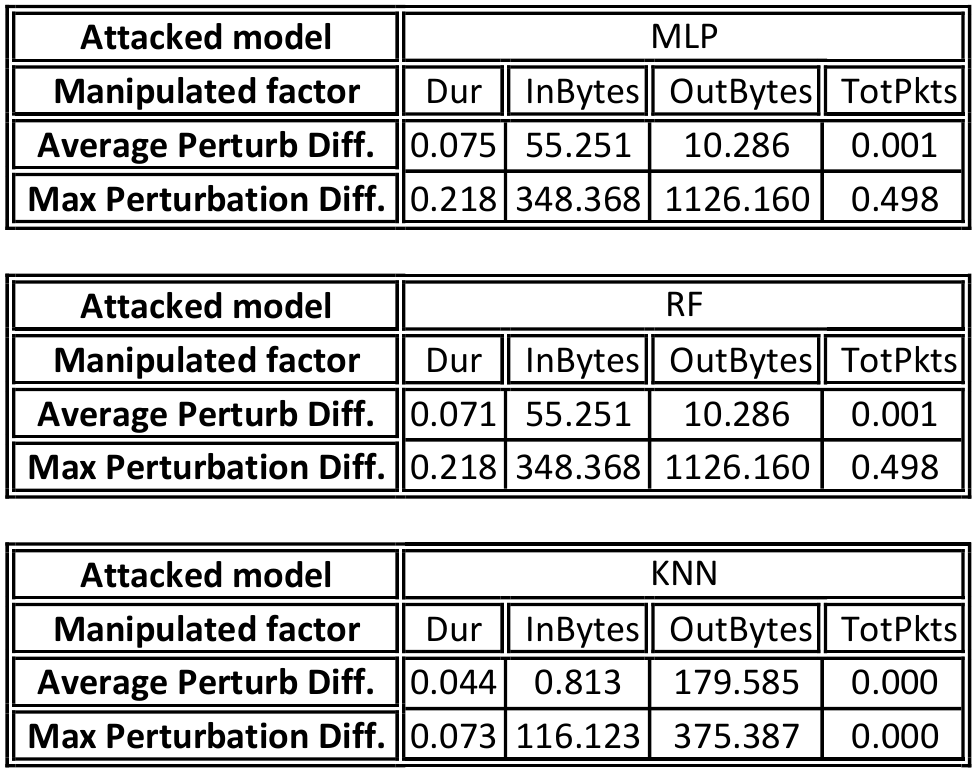}
    \label{tab:perturbation_difference}
\end{table}

After confirming from Table \ref{tab:cicidsstudy} and Table \ref{tab:perturbation_difference}, that the transferability property of adversarial instances allows the attacker to create an adversarial malicious instance to evade the defender's IDS without needing to know its internal architecture while ensuring that the needed perturbations are small enough to be feasible in realistic scenarios, we proceed to create adversarial instances for four botnet attacks, namely Neris, Rbot and Virut from the CTU-13 dataset and Zeus \& Ares from the CSE-CIC-IDS2018 dataset. Note that the attacker and defender models are trained on different data as shown in Figure \ref{fig:dataset_disposition} and have different hyperparameters as shown in  \ref{tab:modelparam}.

Table \ref{tab:adv_ex_datasets_attacker} represents the detection rate of the attacker's models against adversarial botnet attacks. For each botnet type, the attacker generates adversarial instances corresponding to malicious botnets based on the decision boundaries of one of his models using the adversarial generation algorithm \ref{alg:craftadvexalgo}, and then tests these generated adversarial instances on other ML models. On average, we can see that the attacker managed to reduce the detection rate to 15.8\%, 5.3\%, 22.2\% and 21.9\% for Neris, Rbot, Virut and Zeus \& Ares respectively.

\begin{table}[ht]
    \centering
    \caption{Detection rate of the attacker's models against adversarial botnet attacks}
    \includegraphics[width=\columnwidth]{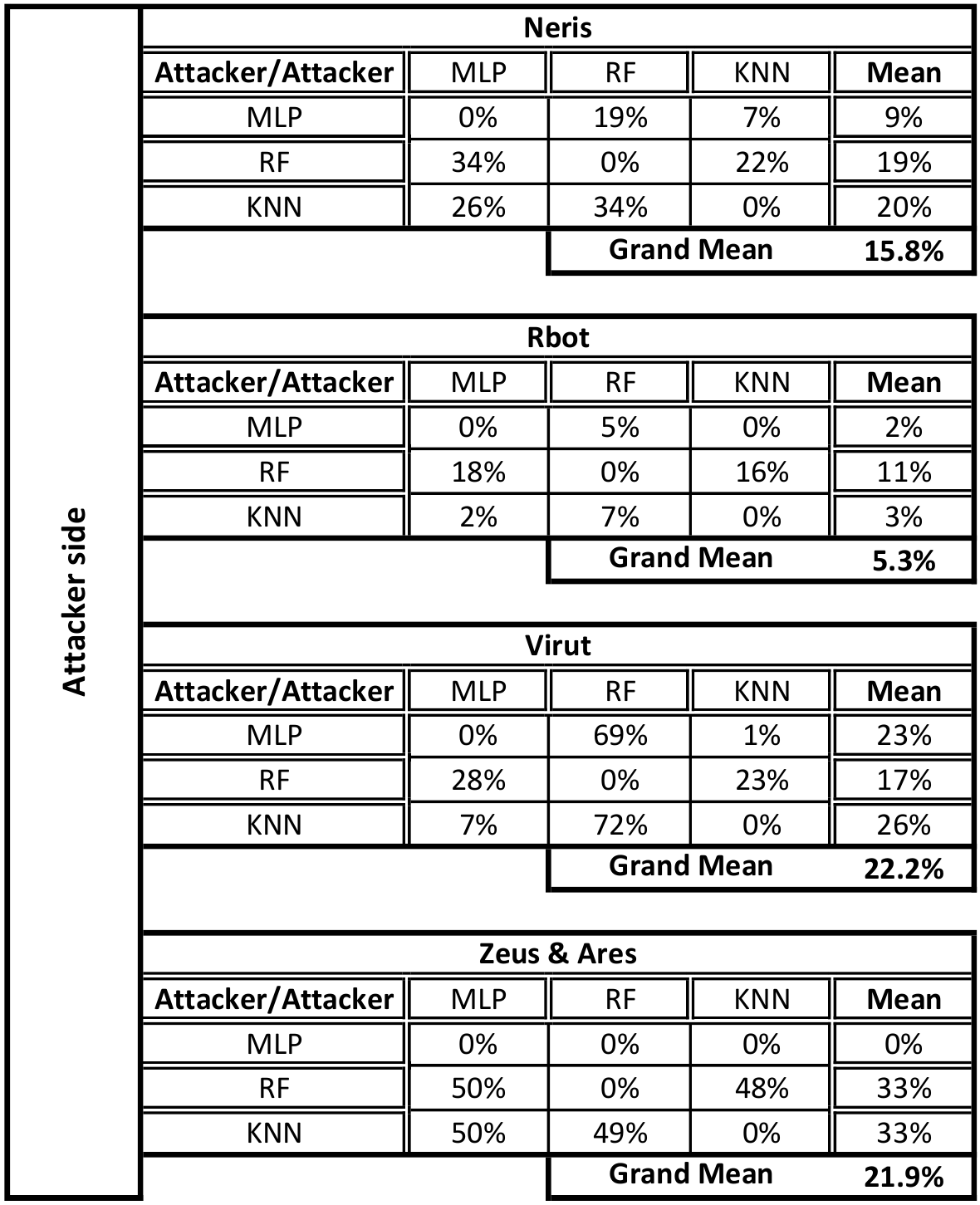}
    \label{tab:adv_ex_datasets_attacker}
\end{table}


On the other hand, Table \ref{tab:adv_ex_datasets_defender} represents the detection rate of the defender's models against adversarial botnet instances generated previously by the attacker. On average, we can see that the attacker managed to reduce the detection rate of the models trained by the defender to 23.0\%, 8.1\%, 37.8\% and 22.4\% for Neris, Rbot, Virut and Zeus \& Ares respectively.  

Using Neris as an example, the attacker generated adversarial instances corresponding to this malware traffic using the decision boundaries of his trained RF model. When testing the effectiveness of his generated adversarial instances on his other trained ML models (i.e., MLP and KNN), he achieves a detection rate of 34\% and 22\%, respectively, as shown in Table \ref{tab:adv_ex_datasets_attacker} and 0\% when tested on RF, since these instances are specifically designed to fool this particular trained model. The attacker then sends these RF-based generated adversarial instances to the defender IDS. The detection rates for the three ML models trained by the defender are 41\%, 19\%, and 23\% for MLP, RF, and KNN, respectively, as shown in Table \ref{tab:adv_ex_datasets_defender}, resulting in an average of 28\% for the defender models. This is a relative success for the attacker since their malicious traffic is only detected 28\% of the time and has almost a three-quarter chance of evading the defender's installed intrusion detection systems in a black-box setting. 

Comparing Table \ref{tab:adv_ex_datasets_attacker} and Table \ref{tab:adv_ex_datasets_defender}, we can observe that the adversarial instances generated by the attacker are, on average, more efficient on his own models than on the models of the defender who uses different training data and hyper-parameters. It can also be observed that intra-transferability has more impact than cross-transferability. Even if this loss is not negligible, the results show good performance on average, both through intra- and cross-transferability.

\begin{table}[ht]
    \centering
    \caption{Detection rate of the defender models against adversarial botnet attacks}
    \includegraphics[width=\columnwidth]{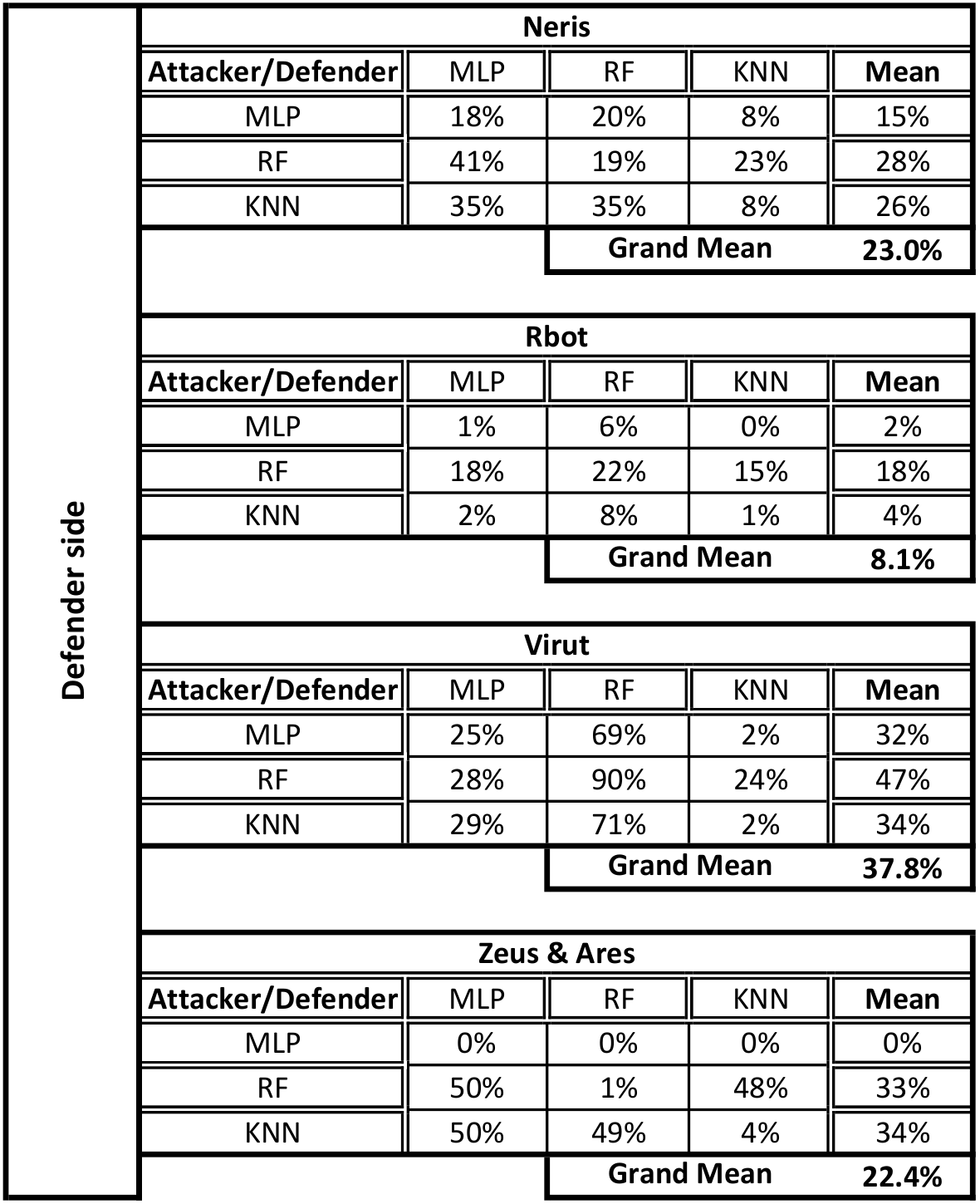}
    \label{tab:adv_ex_datasets_defender}
\end{table}

The time taken to generate an adversarial instance for each of the botnet attacks and models is shown in Table \ref{tab:time_taken_adv_ex}.
It seems that MLP is, for each of the attacks, the algorithm that consumes the most time to generate adversarial instances, followed by RF, which is an ensemble method. On the other hand, KNN seems to be the fastest algorithm to generate adversarial instances for the four botnet attacks, which would make it an interesting option if a trade-off between efficiency and time had to be made.

\begin{table}[ht]

    \centering
    \caption{Time taken (in seconds) to generate 3000 adversarial instances for all botnet attacks using Algorithm \ref{alg:craftadvexalgo}}
    \includegraphics[width=\columnwidth]{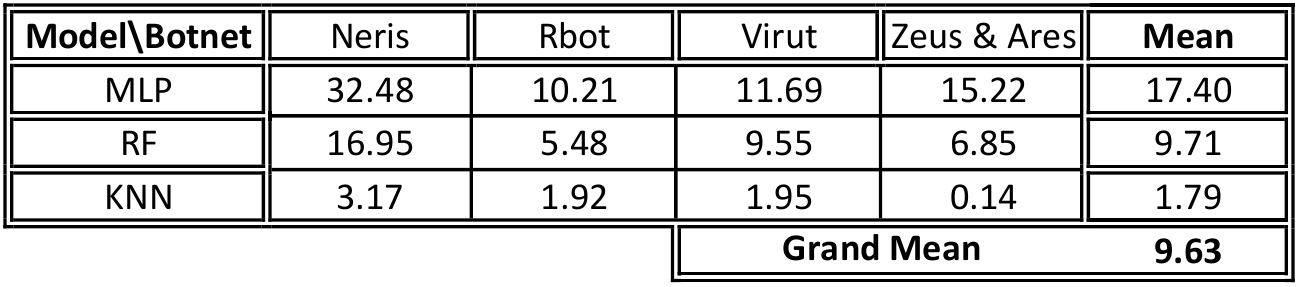}
    \label{tab:time_taken_adv_ex}
\end{table}

\subsection{Proposed Defense effectiveness}

As we discussed in section \ref{suc:AdvPerf}, the attacker is able to evade the defender's NIDS with a high success rate (i.e., a low detection rate) by relying solely on the transferability property of adversarial instances. Our proposed defense aims to negate the effect of the transferability property by adding an adversarial detector to filter out adversarial instances, allowing the NIDS to process only clean traffic.

We first evaluate the performance of the proposed adversarial detector and its sub-detectors by generating adversarial instances based on the CTU-13 and CSE-CIC-IDS2018 datasets. During this analysis, each sub-detector is evaluated as shown in Figure \ref{fig:defprocess}. To measure this performance, various metrics are used: recall as defined in Eq. \ref{eq:recall}; precision as defined in Eq. \ref{eq:precision}; and F1-score as defined in Eq. \ref{eq:f1-score}.

\begin{figure}[ht]
    \centering
    \caption{Performance of our proposed defense against adversarial traffic}
    \includegraphics[width=\columnwidth]{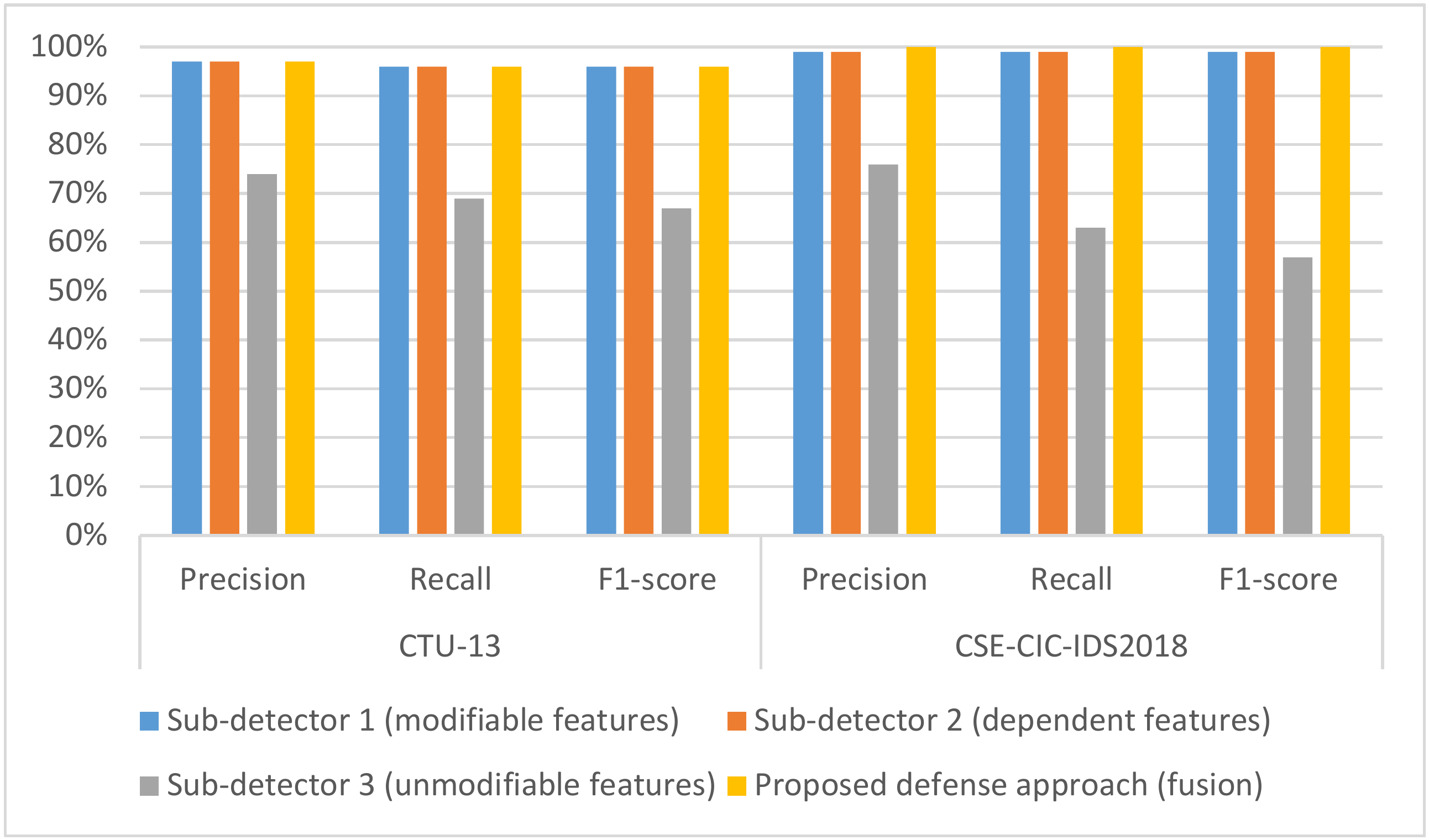}
    \label{fig:initialdefperf}
\end{figure}

As shown in Figure \ref{fig:initialdefperf}, we can see that the first two sub-detectors are performing quite well, reaching more than 96\% for each metric for CTU-13 and 99\% for CSE-CIC-IDS2018, while the last one seems to be less efficient, with performances around 70\%. We can also observe that the final detector performance after Bayesian fusion provides good performances, reaching 97\% for CTU-13 and 100\% for CSE-CIC-IDS2018.


The poorer performance of the third detector can be explained by the fact that it is trained with the group of non-modifiable features. Since the value of these features does not change, it seems that the detector is not able to distinguish adversarial instances from clean instances, thus behaving randomly.

We also note that the fusion of the three detectors slightly improved the overall performance of the proposed defense compared to the individual detectors. The inferior performance of the third detector does not seem to diminish the performance of the proposed defense due to the contextual discounting mechanism, which allows the performance of each individual detector to be taken into account during the fusion stage.

After confirming that our proposed adversarial detector works as intended, we integrate it into our threat scenario. The attacker creates adversarial instances of the four adversarial botnet malware types, namely Neris, Rbot, Virut, and Zeus \& Ares, using the adversarial instance generation algorithm on three ML model decision boundaries, namely MLP, RF, and KNN. The defender, on the other hand, uses an MLP-based NIDS to detect the malicious traffic. In this experiment, attacks are launched twice: the first with the defender's NIDS protected by our proposed adversarial defense and the second without protection. This is done to evaluate the effect of using such a defense mechanism on the performance of the defender's NIDS. The detection rate metric, also known as recall, is used to measure the performance of the NIDS in identifying malicious and benign traffic, but it is also used to assess the performance of the proposed adversarial detector in identifying adversarial and clean traffic. The results are presented in Table \ref{tab:mlpdetectorperfs}, consisting of three sub-tables: 

Table \ref{subtab_a:mlpdetectorperfs} shows the performance results of the defender's MLP-based NIDS against adversarial instances with no adversarial defense. As already shown in Table \ref{tab:adv_ex_datasets_defender}, we see that the attacker has successfully decreased the performance of the defender's NIDS by dropping the average detection rate to 24.8\%.

Table \ref{subtab_b:mlpdetectorperfs} reports the results regarding the ability of our proposed adversarial defense to detect the adversarial instances generated by the attacker. It can be seen that, on average, the proposed adversarial defense was able to detect 93.4\% of the total adversarial botnet traffic sent by the attacker, thus protecting the NIDS from getting evaded by these adversarial instances.

Table \ref{subtab_c:mlpdetectorperfs} represents the detection rate of the defender's NIDS protected by our proposed adversarial defense. In fact, it shows the impact of adversarial instances that have made it through our adversarial detector and reached the NIDS. It can be seen from this table that NIDS has, on average, a detection rate of 96.9\% for any type of machine learning model used for adversarial generation, across all botnet attacks.

These results indicate that NIDS seems to be significantly more robust once the adversarial detection method is used, going from an average detection rate of 21.3\% without defense, as shown in Table \ref{subtab_a:mlpdetectorperfs}, to 96.9\% when using the proposed adversarial detector. This also shows that NIDS is hardly affected by adversarial instances capable of passing the adversarial detector.

\begin{table}[ht]
    \caption{Proposed adversarial defense effectiveness }
    \centering
\begin{subtable}[b]{\columnwidth}
    \centering
    \caption{Detection rate of the defender MLP-based NIDS against adversarial instances without adversarial defense }
    \includegraphics[width=\columnwidth]{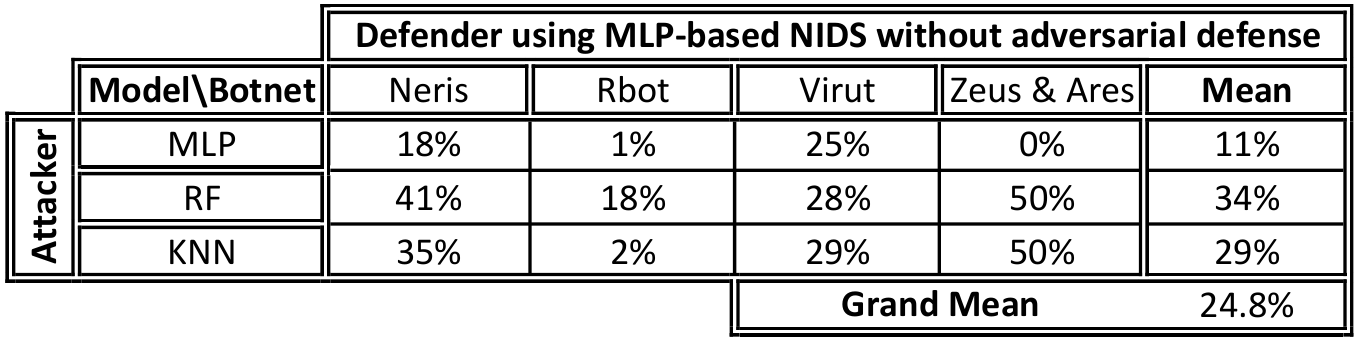}
     \label{subtab_a:mlpdetectorperfs}
\end{subtable}

\begin{subtable}[b]{\columnwidth}
    \centering
    \caption{Detection rate of our proposed adversarial defense}
    \includegraphics[width=\columnwidth]{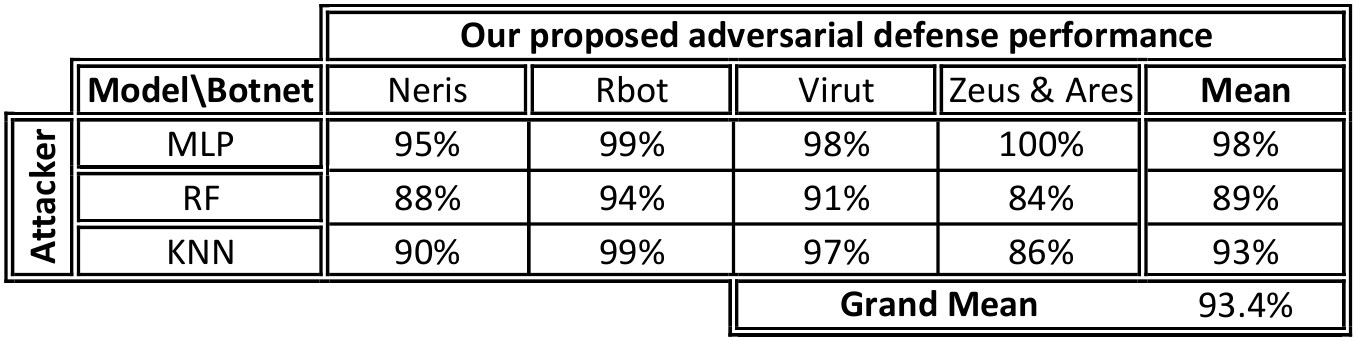}
    \label{subtab_b:mlpdetectorperfs}
\end{subtable}    

\begin{subtable}[b]{\columnwidth}
    \centering
    \caption{Detection rate of the defender MLP-based NIDS against adversarial instances with our proposed adversarial defense}
    \includegraphics[width=\columnwidth]{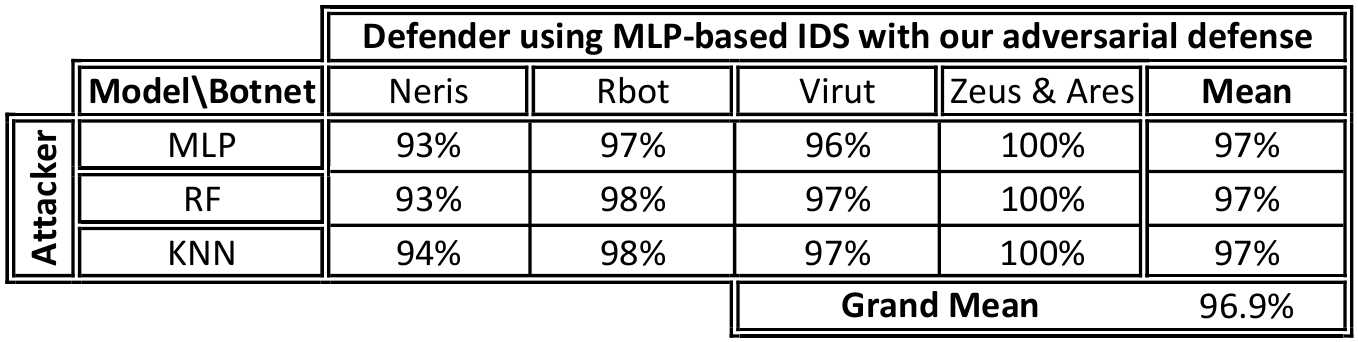}
    \label{subtab_c:mlpdetectorperfs}
\end{subtable}
    
    \label{tab:mlpdetectorperfs}
\end{table}

\subsection{Comparison with existing defensive strategies}
Several countermeasures against the evasion attacks have been proposed in the literature. However, many of them have been shown to be ineffective. Therefore, we decided in this work to compare our proposed defense with adversarial training, which is a state-of-the-art defense that has already demonstrated its effectiveness. To prevent adversarial training from undermining the effectiveness of the IDS in its primary task, we implement this defense in an anomaly detection manner, hence the name adversarial training detection.

 Indeed, adversarial data may be detected by comparing the classification of two models: a baseline model trained solely on non-adversarial samples and a robust adversarial model trained on both non-adversarial and adversarial samples. When applying the two models to a sample, one may assume that if the two models categorize it differently, the sample is adversarial. In principle, if the provided sample is not adversarial, the base model and robust model should correctly identify it. If the sample is adversarial, the base model will categorize it wrong, but the resilient model would classify it properly.
 
 Adversarial training detection defense is built by training a second defense-side MLP-based IDS on data sets with an equal mix of adversarial and non-adversarial samples. We then use the same adversarial instances created by the attacker to attack both models and record their predictions. We infer a final set of predictions by comparing these two sets of predictions so as to classify each instance as adversarial or non-adversarial.

\begin{table}[ht]
    \caption{Detection rate comparison between adversarial training detection and our proposed adversarial defense  }
    \centering

\begin{subtable}[b]{\columnwidth}
    \centering
    \caption{Detection rate of our proposed adversarial defense}
    \includegraphics[width=\columnwidth]{figures/mlpdetectorperf2.pdf}
    \label{subtab_a:mlpdetectorperfs}
\end{subtable}    

\begin{subtable}[b]{\columnwidth}
    \centering
    \caption{Detection rate of adversarial training detection}
    \includegraphics[width=\columnwidth]{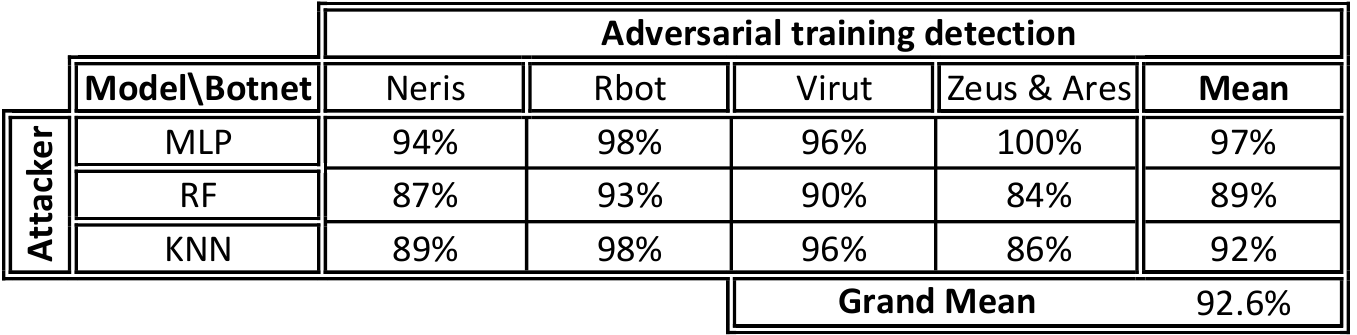}
    \label{subtab_b:otherdefense}
\end{subtable}
    
    \label{tab:mlpdetectorcomp}
\end{table}

In Table \ref{tab:mlpdetectorcomp}, we can see that the achieved results are decent compared to the other state-of-the-art defenses. Our proposed defense slightly outperforms adversarial training detection with an average detection rate of 93.4\% versus 92.6\%. 

Furthermore, compared to other state-of-the-art defenses, our proposed defense is considered a reactive defense, as it does not change the overall performance of the defender's NIDS or general behavior. Therefore, this defense is a better choice when it comes to preserving the general performance of the model, which is not the case with adversarial training, for example, which reduces the performance of the underlying model.


\section{Conclusion}
\label{conclusion}
The potential of using NIDS based on machine learning algorithms raises intriguing security issues. Indeed, despite their impressive performance, these ML models are prone to various types of adversarial attacks, especially evasion attacks. Due to their prevalence and feasibility among all adversarial attacks, evasion attacks were considered in this paper to generate botnet adversarial attacks capable of evading the intrusion detection system. 

The proposed framework includes two main contributions. The first is a realistic adversarial algorithm capable of generating valid adversarial network traffic by adding small perturbations, thus evading NIDS protection with high probability while maintaining the underlying logic of the botnet attack. To the best of our knowledge, this is the first complete black-box botnet attack that proposes to evade NIDS by exploiting the transferability property, and without using any query method, with very limited knowledge of the target NIDS, which acts on the traffic space, while respecting the domain constraints.

The second component of the proposed framework is a reactive defense that limits the impact of the proposed attack. This defense, inspired by adversarial detection, capitalizes on the fact that it does not change the initial performance of the NIDS since it provides an additional layer of security independent of the model. The proposed defense is considered modular because it uses an ensemble method called bagging yet can use any type of machine learning algorithm. In addition to this ensemble method, it also includes a contextual discounting method that improves the overall performance of the defense.

The results showed that the proposed defense is able to detect most adversarial botnet traffic, showing promising results with respect to state-of-the-art defenses. Since the proposed framework is easily adaptable to other domains, evaluating its performance in other highly constrained domains would be an interesting future work.

\section*{Declaration of competing interest}
The authors declare that they have no known competing financial interests or personal relationships that could have appeared to influence the work reported in this paper.

\section*{CRediT authorship contribution statement}
\textbf{Islam Debicha:} Conceptualization, Software, Validation, Methodology, Writing - Original Draft. \textbf{Benjamin Cochez:} Software, Visualization, Writing - Original Draft. \textbf{Tayeb Kenaza:} Validation, Writing - Review \& Editing, Supervision. \textbf{Thibault Debatty:} Validation, Writing - Review \& Editing, Supervision. \textbf{Jean-Michel Dricot:} Validation, Writing - Review \& Editing, Supervision. \textbf{Wim Mees:} Funding acquisition, Supervision.

\bibliography{mybibfile}

\bigskip 

\piccaptioninside

\parpic[l]{\includegraphics[width=0.3\linewidth, height=0.2\textwidth]{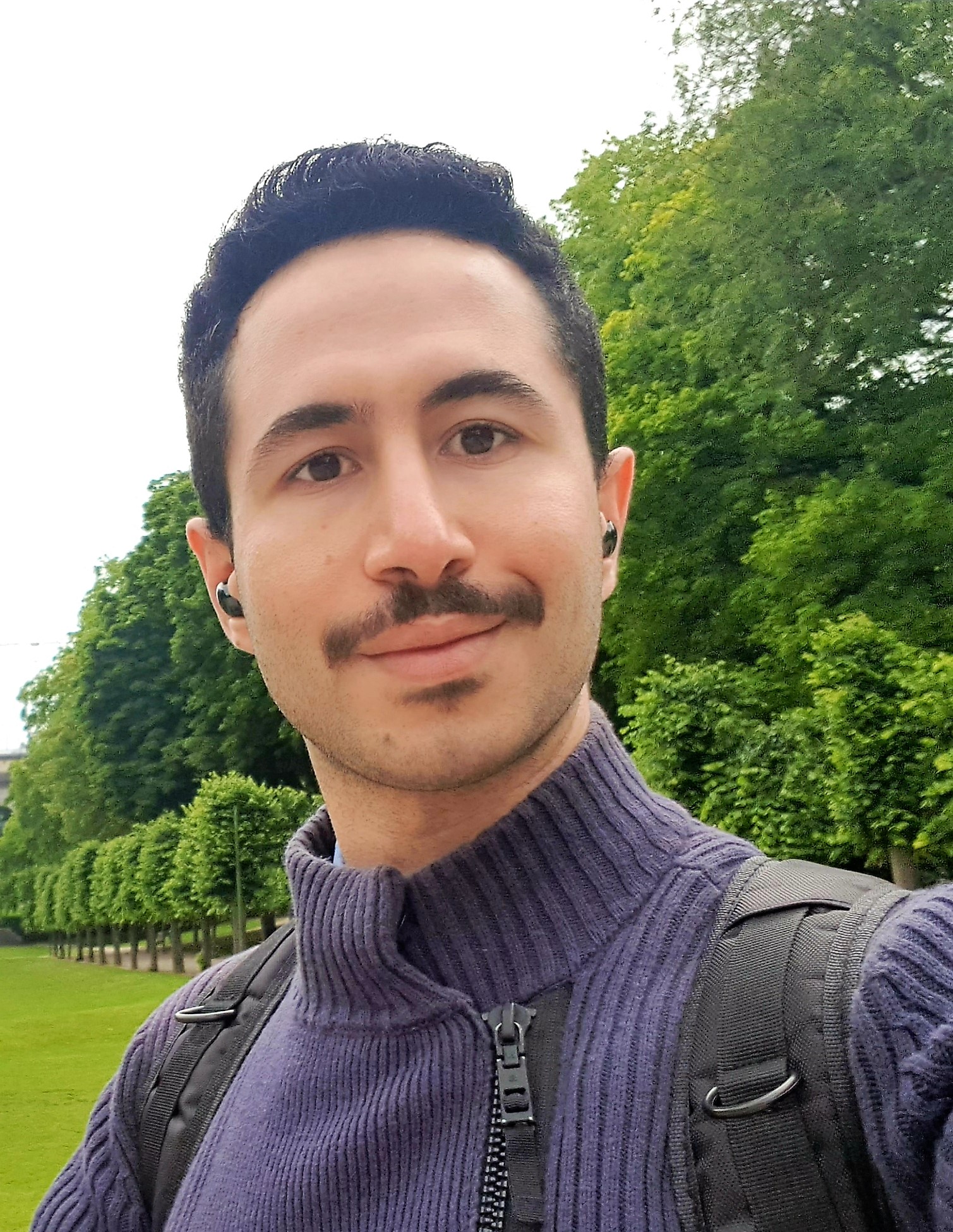}}

\textbf{Islam Debicha} received his master's degree in Computer Science with a focus in network security in 2018. He is pursuing a joint Ph.D. in machine learning-based intrusion detection systems at ULB and ERM, Brussels, Belgium. He works at ULB cybersecurity Research Center and Cyber Defence Lab on evasion attacks against machine learning-based intrusion detection systems. He is the author or co-author of 6 peer-reviewed scientific publications. His research interests include defenses against adversarial examples, machine learning applications in cybersecurity, data fusion, and network security.

\piccaptioninside

\parpic[l]{\includegraphics[width=0.3\linewidth, height=0.2\textwidth]{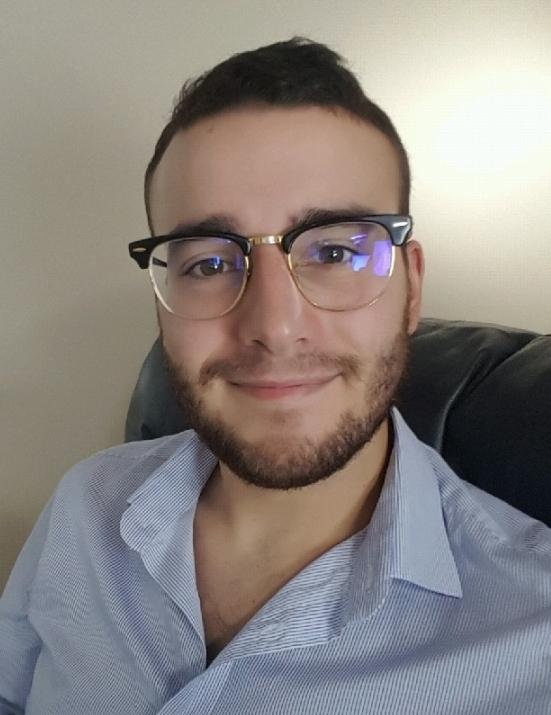}}
\textbf{Benjamin Cochez} received his master's degree in Cybersecurity from the Université Libre de Bruxelles (ULB). He works as a cybersecurity consultant for a consultancy company based in Brussels, and his main areas of interest are cloud security, endpoint security, machine learning and adversarial learning. \\

\piccaptioninside

\parpic[l]{\includegraphics[width=0.3\linewidth, height=0.2\textwidth]{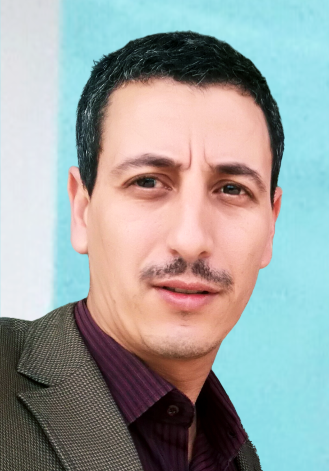}}
\textbf{Tayeb Kenaza }received a Ph.D. degree in computer science from Artois University, France, in 2011. Since 2017, he is the Head of the Computer Security Laboratory at the Military Polytechnic School of Algiers. He is currently a full professor in Computer Science Department at the same school. His research and publication interests include Computer Network Security, Wireless Communication Security, Intelligent Systems, and Data Mining.

\piccaptioninside

\parpic[l]{\includegraphics[width=0.3\linewidth, height=0.2\textwidth]{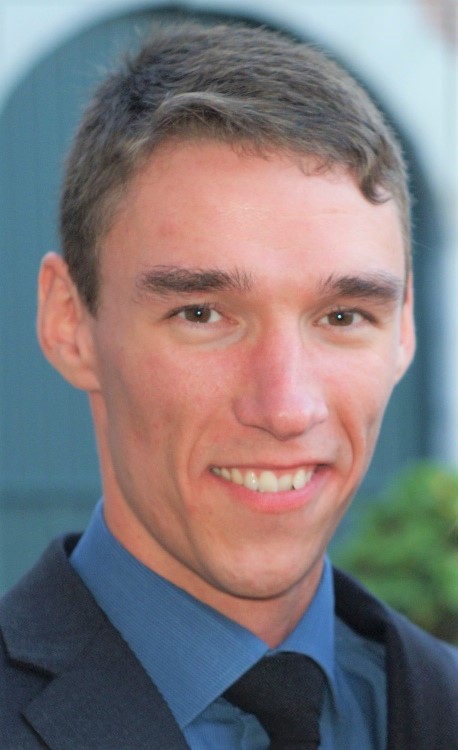}}
\textbf{Thibault Debatty} obtained a master's degree in applied engineering sciences at the Royal Military Academy (RMA) in Belgium, followed by a master's degree in applied computer science at the Vrije Universiteit Brussel (VUB). Finally, he obtained a Ph.D. with a specialization in distributed computing at both Telecom Paris and the RMA. He is now an associate professor at the RMA, where he teaches courses in networking, distributed information systems, and information security. He is also president of the jury of the Master of Science in Cybersecurity organized by the Université Libre de Bruxelles (ULB), the RMA, and four other institutions.

\piccaptioninside

\parpic[l]{\includegraphics[width=0.3\linewidth, height=0.2\textwidth]{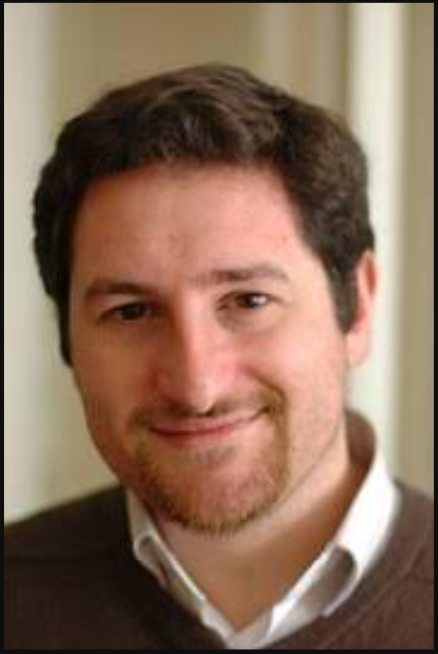}}
\textbf{Jean-Michel Dricot} leads research on network security with a specific focus on the IoT and wireless networks. He teaches communication networks, mobile networks, the internet of things, and network security. Prior to his tenure at the ULB, Jean-Michel Dricot obtained a Ph.D. in network engineering, with a focus on wireless sensor network protocols and architectures. In 2010, Jean-Michel Dricot was appointed professor at the Université Libre de Bruxelles, with tenure in mobile and wireless networks. He is the author or co-author of more than 100+ papers published in peer-reviewed international Journals and Conferences and served as a reviewer for European projects.

\piccaptioninside

\parpic[l]{\includegraphics[width=0.3\linewidth, height=0.2\textwidth]{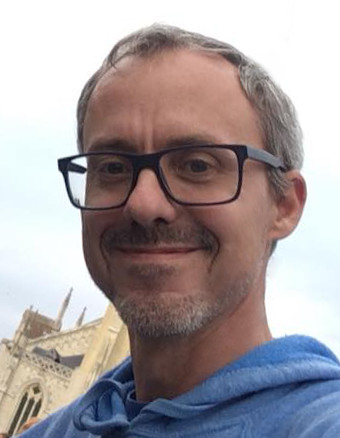}}
\textbf{Wim Mees} is Professor in computer science and cyber security at the Royal Military Academy and is leading the Cyber Defence Lab. He is also teaching in the Belgian inter-university Master in Cybersecurity, and in the Master in Enterprise Architecture at the IC Institute. Wim has participated in and coordinated numerous national and European research projects as well EDA and NATO projects and task groups. 
\end{document}